\documentclass[aps, pre, a4paper, twocolumn]{revtex4-2}
\usepackage{graphicx} 
\usepackage{enumitem}
\usepackage{float}
\usepackage{amsmath, amssymb, amsfonts}
\usepackage{hyperref} 
\hypersetup{
colorlinks = true, 
linkcolor = blue
}
\usepackage{newtxtext, newtxmath}

\usepackage{microtype}
\begin{document}
\title{Canonical equilibrium of mean-field $O(n)$~models in presence of random fields}
\author{Soumya Kanti Pal}
\email{Corresponding author; Email: soumya.pal@tifr.res.in}
\author{Shamik Gupta}
\affiliation{Department of Theoretical Physics, Tata Institute of Fundamental Research, Homi Bhabha Road, Mumbai 400005, India}
\date{\today}
\begin{abstract}
We study canonical-equilibrium properties of Random
Field $O(n)$ Models involving classical continuous vector spins of $n$ components with mean-field interactions and subject to
disordered fields acting on individual spins. To this end, we employ two complementary approaches: the mean-field approximation, valid for any disorder distribution, and the replica trick, applicable when the disordered fields are sampled from a Gaussian distribution. On the basis of an exact analysis, we demonstrate that when replica symmetry holds, both the approaches yield identical expression for the free energy per spin of the system. As consequences, we study the case of $n=2$ ($XY$ spins) and that of $n=3$ (Heisenberg spins) for two representative choices of the disorder distribution, namely, a Gaussian and a symmetric bimodal distribution. For both $n=2$ and $n=3$, we demonstrate that while the magnetization exhibits a continuous phase transition as a function of temperature for the Gaussian case, the transition could be either continuous or first-order with an emergent tricriticality when the disorder distribution is bimodal. We also discuss in the context of our models the issue of self-averaging of extensive variables near the critical point of a continuous phase transition.
\end{abstract}
\maketitle

\section{Introduction}
\label{sec:model}
Quenched disorder is known to affect static and dynamic properties of many-body interacting systems both in and out of equilibrium. Systems with quenched disorder when in equilibrium host many intriguing phenomena such as occurrence of the spin-glass phase~\cite{Anderson_1975,SK,Binder_young_review,Parisi_infinite_solutions}, weak-ergodicity breaking~\cite{weak-ergodicity}, Griffith's singularities~\cite{Griffith_singularities}, anomalous relaxation~\cite{Natterman_relaxation,Randeria}, absence of self-averaging near a critical point~\cite{Aharony_Harris_self_averaging,Wiseman_self_averaging}, etc. In systems driven out of equilibrium, quenched disorder manifests itself in several remarkable phenomena in non-equilibrium steady states of such systems, e.g., phase separation~\cite{krug_phase_separation}, non-trivial correlation functions~\cite{Tripathy_Barma_correlation}, appearance of shocks~\cite{Kirone_mallick_shocks}, localization effects~\cite{Stinchcombe_localization}, and many others. 

A paradigmatic model to study the effects of disorder on equilibrium properties of finite-dimensional magnetic systems has been the well-known random-field Ising model (RFIM) involving Ising spins with nearest-neighbor ferromagnetic coupling on a lattice ~\cite{Imry-Ma-RFIM}, in which the magnetic field at every lattice site is a quenched-disordered random variable. It was argued that there is a critical lattice dimension $d_c$ below which even an infinitesimal random field is sufficient to destabilize any long-range order in the system. Subsequently, it was shown that at the critical point, a $d$-dimensional system with quenched-disordered fields is equivalent to a $(d-2)$-dimensional system without random fields, provided the dimension $d$ exceeds a certain upper critical dimension~\cite{Parisi_sourlas_supersymmetry}. For the RFIM, extracting phase transition points and critical exponents has been non-trivial and is still an area with many open questions to be resolved \cite{Fytas3,Fytas2,Fytas1}. The RFIM with mean-field coupling was first analyzed in Ref.~\cite{Aharony}, for which renormalization group techniques were used to show that the phase diagram for symmetric disorder distributions with a minimum at zero exhibits a tricritical point separating continuous from first-order transitions. The generalization of this result to continuous vector spins with 
$n$-components (i.e., $O(n)$ spins, see Eq.~\eqref{eq:O(n) spins} below) is an active area of research in the field of statistical physics of disordered systems. Some recent work in this area for $n=2$ (i.e., for $XY$ spins) include consideration of random crystal fields~\cite{Sumedha-Barma-crystal-field,Rajiv-Barma}, random orientation fields~\cite{Sumedha-Barma-LDT, lupo,PhysRevB.110.184205,del2024most,Rajiv-Barma}, Markov random fields~\cite{markov-random-field}, all in the mean-field limit, as also demonstration of replica-symmetry-breaking in the two-dimensional $XY$ model with onsite disordered fields~\cite{Cardy,P.Doussal-Thierry}. Across these studies, the central issues of investigation have been the following: (i) How does the phase transition of generic $O(n)$ spin systems with quenched-disordered fields get affected by the presence of disorder in relation to the corresponding system without disorder? (ii) In this regard, what is the essential role played by (a) the specific value of $n$, and (b) the choice of the disorder distribution? These questions constitute the primary motivation of the present work.

The models we consider here form the class of Random Field $O(n)$ Models, which involves classical continuous vector spins of $n$ components with mean-field interactions and subject to disordered fields acting on individual spins. The Hamiltonian reads as 
\begin{equation}\label{eq:H}
\mathcal{H}= -\frac{J}{2N} \sum_{i,j=1}^N \Vec{S}_i \cdot \Vec{S}_j -\sum_{i=1}^N \Vec{h}_i \cdot \Vec{S}_i,
\end{equation}
where the ferromagnetic coupling $J>0$ is the same for every pair of spins. The spin vector $\Vec{S}_i$ has $n$ components $(S_i^{(1)}, S_i^{(2)},\ldots, S_i^{(n)})$, and is of unit length: $\sum_{\alpha=1}^n (S_i^{(\alpha)})^2 = 1~\forall~i$. We follow the convention that Greek letters $\alpha,\gamma$, etc. denote spin components, while Roman letters $i,j,k,$ etc. denote spin indices and Roman letters $a,b,c,$ etc. stand for replica indices. One can express the spin components in $n$-sphere polar coordinates as 
\begin{align}\label{eq:O(n) spins}
&S_i^{(1)}= \cos{\theta_i^{(1)}}, \nonumber \\
&S_i^{(2)}=\sin{\theta_i^{(1)}} \cos{\theta_i^{(2)}}, \nonumber\\
&S_i^{(3)}=\sin{\theta_i^{(1)}} \sin{\theta_i^{(2)}} \cos{\theta_i^{(3)}}, \nonumber\\
& \vdots \nonumber\\
&S_i^{(n-1)}= \sin{\theta_i^{(1)}} \sin{\theta_i^{(2)}} \ldots  \sin{\theta_i^{(n-2)}} \cos{\theta_i^{(n-1)}}, \nonumber\\
&S_i^{(n)}= \sin{\theta_i^{(1)}} \sin{\theta_i^{(2)}} \ldots  \sin{\theta_i^{(n-2)}} \sin{\theta_i^{(n-1)}}, \nonumber\\
\end{align}
with the polar angles in the range $0 \leq \theta_i^{(1)},\theta_i^{(2)}, \ldots \theta_i^{(n-2)} \leq \pi$ and $0 \leq \theta_i^{(n-1)} <2\pi$.

In Eq.~\eqref{eq:H}, each of the $n$ components of the fields $\Vec{h}_i$ for every $i$ is chosen independently from a given distribution $P(h)$ with mean zero and variance $\sigma>0$. Consequently, the $nN$ variables $\{h_i^{(1)},h_i^{(2)}, \ldots,h_i^{(n)}\}_{1\le i \le N}$ constitute a set of quenched-disordered random variables.We consider in this study two representative examples of $P(h)$, one of which is unimodal while the other is bimodal, namely, a Gaussian and a bimodal distribution, given respectively by
\begin{align}
&P(h)=\frac{1}{\sqrt{2\pi\sigma^2}} \exp{\left(-\frac{h^2}{2\sigma^2}\right)}, \label{eq:Gaussian}\\
&P(h)=\frac{1}{2 }[\delta(h-h_0) + \delta(h+h_0)]\label{eq:bimodal}.
\end{align}
Here, $\sigma>0$ and $h_0>0$ are given parameters that may be taken to denote the strength of disorder in each of the two cases of the disorder distribution. In the case of the bimodal distribution, the parameter $h_0$ sets the variance of the distribution, which equals $h_0^2$; the same for the Gaussian distribution equals $\sigma^2$. 

In this work, we study the properties of the system~\eqref{eq:H} in canonical equilibrium at temperature $T$. To this end, we pursue two complementary approaches. The first one is that of the mean-field (MF) approximation, which is an obvious choice given the all-to-all interaction between the spins in system~\eqref{eq:H}. However, intricacies arise in applying a naive mean-field approximation owing to the presence of the quenched-disordered fields in the system. Our modified MF approach involves two levels of approximation. Firstly, we consider the system for a fixed realization of the disordered fields and study it in terms of a mean-field Hamiltonian expressed in terms of the thermal-averaged magnetization. Here, the underlying assumption is that the thermal average of every spin component fluctuates only negligibly from one spin to another, despite the fact that each spin on its own is evolving in a different environment owing to the presence of the disordered fields varying from spin to spin. The next level of approximation is to assume that the thermal-averaged magnetization, which in principle varies from one disorder realization to another, nevertheless takes up the same value across disorder realizations (self-averaging property). This is expected to be valid provided the disorder distribution does not have fat tails, which is the case for the two representative choices in Eqs.~\eqref{eq:Gaussian} and~\eqref{eq:bimodal}. The outcome of our analysis is an explicit expression for the disorder-averaged free energy per spin, from which ordering properties of the system in equilibrium and in particular information about phase transitions in equilibrium can be straightforwardly derived.

In the second complementary approach, we invoke for the case of Gaussian distribution for the disordered fields, the celebrated replica trick commonly used to analyze systems with quenched disorder, especially, spin-glass systems. In computing the disorder-averaged free energy per spin, one is required to compute the disorder-averaged canonical partition function 
 $\overline{\ln Z}$. Such a computation within the replica trick is reduced to computing the disorder average of the product of the partition function for $p$ replicas, $\overline{Z^p}$, with $p$ an integer, where each replica is defined as an identical copy of the system with identical disorder realizations but with dynamics leading to potentially-different equilibrium configurations. The nontriviality of the approach lies in recovering $\overline{\ln Z}$ from the behavior of $(\overline{Z^p}-1)/p$ as $p$ approaches continuously to zero. Under the assumption of replica symmetry, which asserts that the equilibrium attained across the replicas is the same, we show on the basis of our exact computation that results from the MF and the replica approach coincide exactly.

 In the following part of the paper, we use our exact expression for the disorder-averaged free energy per spin to derive phase transition behavior for two representative systems in the class of Random
Field $O(n)$ Models, namely, for $n=2$ ($XY$ spins) and $n=3$ (Heisenberg spins), considering for each the two representative disorder distributions, Eqs.~\eqref{eq:bimodal} and~\eqref{eq:Gaussian}. We report universal behavior across the studied models, namely, that both for $n=2$ and $n=3$ and with the Gaussian distribution for the disordered fields, the system shows in equilibrium (i) at a non-zero temperature a continuous phase transition between a low-$T$ ferromagnetic and a high-$T$ paramagnetic phase, and (ii) at $T=0$ a continuous phase transition between a low-$\sigma$ ferromagnetic and a high-$\sigma$ paramagnetic phase. Contrarily and remarkably, when the distribution is bimodal, both the models when in equilibrium at a non-zero temperature show either a first-order or a continuous phase transition in the $(h_0,T)$-plane, with the two transition lines meeting at a tricritical point. The $T=0$ transition as a function of $h_0$ is however first-order.   

The paper is organized as follows. In Section~\ref{sec:equilibrium}, we discuss the equilibrium properties of the system, considering in turn the two complementary approaches, the mean-field approximation, Section~\ref{sec:MF}, and the replica-trick approach~\ref{sec:replica}. Sections~\ref{sec:n2} and~\ref{sec:n3} are devoted to respectively derivation of the phase-transition behavior of the $n=2$ and the $n=3$ case. In Section~\ref{sec:absence of self-averaging}, we discuss in the context of our models the issue of self-averaging of extensive variables near the critical point of a continuous phase transition. The paper ends with conclusions in Section~\ref{sec:conclusion}.

\section{Equilibrium properties}
\label{sec:equilibrium}
Here, we investigate the properties of the system (\ref{eq:H}) in the thermodynamic limit $N\to \infty$ and in canonical equilibrium at temperature $T=1/\beta$ [we work in units in which the Boltzmann constant is unity]. Although a mean-field model,  the local environment of each spin is not identical owing to the disordered fields $\Vec{h}_i$. The exactness of mean-field theory in cases of quenched-disordered fields and couplings has been studied in Refs.~\cite{Tsuda} and~\cite{Mori}, respectively. The main problem in writing down a free-energy expression for a quenched-disordered Hamiltonian relates to making a distinction between computing the logarithm of disorder-averaged partition function, $\log\overline{\mathcal{Z}}$, and the disorder-averaged logarithm of the partition function, $\overline{\log \mathcal{Z}}$. It is well known that the former is unphysical, while the latter is physical, based on the self-averaging property of the free energy (for further discussions, see~Refs.~\cite{dotsenko,Castellani,Menon}).

\subsection{Mean-field approximation}
\label{sec:MF}
In order to study the canonical-equilibrium properties of system~\eqref{eq:H}, we develop in this part an effective mean-field (MF) approach, which involves two levels of approximation: (i) For a fixed realization of disordered fields, we study the system in terms of a mean-field Hamiltonian expressed in terms of the thermal-averaged magnetization (averaging with respect to the canonical-equilibrium measure);  (ii) Next, we invoke that the thermal-averaged magnetization is self-averaging in the limit $N\to \infty$, in that it has the same value across different disorder realizations. Let us now proceed to construct first the mean-field Hamiltonian.

\subsubsection{Constructing the mean-field Hamiltonian}
To proceed, let us denote the mean-field Hamiltonian by $\mathcal{H}_\mathrm{MF}(\{\Vec{h}_i\})$ for a fixed realization of disordered fields $\{\Vec{h}_i\}$. Clearly, there are $n$ components of the thermal-averaged magnetization vector  $\Vec{\Tilde{m}}=(\Tilde{m}^{(1)},\Tilde{m}^{(2)}, \ldots, \Tilde{m}^{(n)})$, with $\Tilde{m}^{(\alpha)}=(1/N)\sum_{i=1}^N \Tilde{m}_i^{(\alpha)}$ and $\Tilde{m}_i^{(\alpha)} \equiv \langle S_i^{(\alpha)}\rangle$, where angular brackets denote thermal averaging. The norm of this vector, given by $\Tilde{m} = |\Vec{\Tilde{m}}|=\sqrt{\sum_{\alpha=1}^n (\Tilde{m}^{(\alpha)})^2}$, serves as the order parameter of the system. Notice that there is no a priori requirement for $\Tilde{m}_i^{(\alpha)}$ to be independent of disorder realizations and be only dependent on $\beta$. 

We now write the component of any spin in terms of its thermal-averaged realization-dependent value $\Tilde{m}_i^{(\alpha)}$ as $ S_i^{(\alpha)}=\Tilde{m}_i^{(\alpha)}  + \delta S_i^{(\alpha)}$, where $\delta S_i^{(\alpha)}$ denotes deviation from the thermal average. In the first step of approximation, we assume that the thermal average is the same for every spin, despite the fact that every spin has got different local environment: $\Tilde{m}_i^{(\alpha)}=\Tilde{m}^{(\alpha)}$ for all $i=1,2,\ldots,N$ and $\alpha=1,2,\ldots,n$.  The above steps result in the expression 
\begin{align}
S_i^{(\alpha)}S_j^{(\alpha)}&=(\Tilde{m}^{(\alpha)})^2 + \Tilde{m}^{(\alpha)} (\delta S_i^{(\alpha)} + \delta S_j^{(\alpha)})+ \delta S_i^{(\alpha)} \delta S_j^{(\alpha)}.
\end{align}
Next, retaining up to the linear term and neglecting higher-order fluctuations, i.e., the term $\delta S_i^{(\alpha)} \delta S_j^{(\alpha)}$, we get 
\begin{align}
\label{eq4}
S_i^{(\alpha)} S_j^{(\alpha)}& =(\Tilde{m}^{(\alpha)})^2+m^{(\alpha)} (S_i^{(\alpha)} + S_j^{(\alpha)} -2m^{(\alpha)})\nonumber \\
&=m^{(\alpha)} (S_i^{(\alpha)} + S_j^{(\alpha)} -m^{(\alpha)}),
\end{align}
which used in Eq.~\eqref{eq:H} finally leads to the following mean-field Hamiltonian:
\begin{align}
\label{eq5}
\mathcal{H}_\mathrm{MF}=\frac{N J \Tilde{m}^2}{2} - \sum_{i=1}^N \Vec{S}_i \cdot (\Vec{h}_i + J\Vec{\Tilde{m}}). 
\end{align}

For future purposes, let us define the quantity 
\begin{align}
\Vec{K}_i \equiv \Vec{h}_i + J\Vec{\Tilde{m}}, 
\label{eq:Ki}
\end{align}
which allows to rewrite Eq.~\eqref{eq5}  as 
\begin{align}
\mathcal{H}_\mathrm{MF}=\frac{N J \Tilde{m}^2}{2} - \sum_{i=1}^N \Vec{S}_i \cdot \Vec{K}_i.
\end{align}
The thermal average computed with respect to the above mean-field Hamiltonian gives the following self-consistent relation for $\Tilde{m}^{(\alpha)}$:
\begin{align}
{\langle S_i^{(\alpha)} \rangle}=\Tilde{m}_i^{(\alpha)}=\Tilde{m}^{(\alpha)}=\frac{1}{z_i}\int d\Vec{S}_i~ S_i^{(\alpha)} \exp[\beta \Vec{S}_i \cdot \Vec{K}_i],
\end{align}
where, with the single-spin integral $\int d \Vec{S}_i$ defined as 
\begin{align}
\int d \Vec{S}_i\equiv \int_{0}^{\pi} d\theta^{(1)}   \ldots \int_{0}^{2\pi} d\theta^{(n-1)} ~ \mathcal{J}(\theta^{(1)}, \ldots, \theta^{(n-1)}),   
\end{align}
and $\mathcal{J}(\theta_i^{(1)}, \ldots, \theta_i^{(n-1)})$ denoting the unit $n$-sphere volume element, we have the single-spin partition function
\begin{align}
z_i&\equiv\int d\Vec{S}_i ~\exp[\beta \Vec{S}_i \cdot \Vec{K}_i]. 
\end{align}
Having obtained the mean-field Hamiltonian as in Eq.~\eqref{eq5}, we now move on to obtain the mean-field partition function.

  \subsubsection{Computing the mean-field partition function}
The mean-field partition function reads as 
\begin{align}
{\label{eq6}}
\mathcal{Z}_\mathrm{MF}&=\exp{\left(-\frac{\beta N J \Tilde{m}^2}{2}\right)} \int \prod_{j=1}^N d \Vec{S}_j~\exp{\left(\beta \sum_{i=1}^N \Vec{S}_i \cdot \Vec{K}_i\right)}\nonumber \\
&=\exp{\left(-\frac{\beta N J \Tilde{m}^2}{2}\right)}\prod_{j=1}^N z_j,
\end{align} 
yielding 
\begin{align}
{\label{eq7}}
\log \mathcal{Z}_\mathrm{MF} = - \frac{\beta N J \Tilde{m}^2}{2} +\sum_{j=1}^N\log{z_j}. 
\end{align}
Now, performing the average over all possible realizations of the disordered fields $\{h_{i}^{(1)},h_{i}^{(2)}, \ldots,h_{i}^{(n)}\}_{1\le i \le N}$ in Eq.~\eqref{eq7}, we obtain  
\begin{align}{\label{eq8}}
\overline{\log \mathcal{Z}_\mathrm{MF} }&=-\frac{\beta N J }{2}\int \prod_{i=1}^N d\Vec{h}_i~ P(\Vec{h}_i) \Tilde{m}^2 \nonumber 
\\&~~~~ + \int \prod_{i=1}^N d\Vec{h}_i~ P(\Vec{h}_i) \sum_{j=1}^N\log{z_j},
\end{align}
where we have
\begin{align}
 P(\Vec{h}_i)d\Vec{h}_i=\prod_{\alpha=1}^n P(h_i^{(\alpha)})dh_i^{(\alpha)},
 \end{align}
using the fact that for every spin, the components of the disordered field are sampled independently from the common distribution $P(h)$.
We now aim to obtain the disorder-averaged free energy of the system.

\subsubsection{Obtaining the disorder-averaged free energy}
At this point, we invoke the self-averaging property of the thermal-averaged magnetization appearing in Eq.~\eqref{eq8}. This implies that the distribution of any component of thermal-averaged magnetization across different disorder realizations is sharply peaked, indicating negligible sample-to-sample fluctuations. Such a property is expected to hold as long as the disorder distribution is not anomalously broad or pathological. The self-averaging property motivates us to replace $\Tilde{m}^{(\alpha)}$ with the corresponding disordered-averaged quantity $m^{(\alpha)}$, given by the integral $\int  \prod_{i=1}^N d\Vec{h}_i P(\Vec{h}_i)  ~ \Tilde{m}^{(\alpha)}$. This forms the second step of our mean-field approximation, and which on using Eq.~\eqref{eq8} directly leads us to the final expression for the disorder-averaged free energy per spin as
\begin{align} 
\label{eq_F}
\frac{\mathcal{F}}{N}& = -k_BT\, \frac{\overline{\log \mathcal{Z}_\mathrm{MF} }}{N}\nonumber \\
&=\frac{Jm^2}{2} -   \frac{1}{\beta}\int d{\Vec{h}}~ P(\Vec{h})   \log\left\{\int d \Vec{S} \exp[\beta \Vec{S} \cdot (\Vec{h} + J\Vec{m})]   \right\}.   
\end{align}

\subsubsection{On the validity of the mean-field approximation}
We conclude this part by stressing the fact that the derivation of the mean-field-approximated free energy per spin in Eq.~\eqref{eq_F} crucially relies on the condition that the underlying disorder distribution has sharply-decaying tails; however, the issue of how sharp the tails should be remains unsettled. Namely, for a Pareto distribution given by 
\begin{align}
\label{eq:Pareto}
P(h) \propto \frac{1}{h_0 + h^{2\delta}};~-\infty <h<\infty,
\end{align}   
one does not know exactly the threshold exponent $\delta$ above which this mean-field approximation would work. In our study, the representative distributions we consider have at most exponentially-decaying tails, and therefore, the mean-field approximation is expected to be valid for them.

In the next part of the paper, we consider the random fields $\{\vec{h}_i\}$ to follow the Gaussian distribution~\eqref{eq:Gaussian} and employ the celebrated replica trick as an alternative to the mean-field approximation discussed above to obtain the disorder-averaged free energy of the system. We will demonstrate that for the case of the Gaussian distribution, one obtains using the two approaches identical expression for the disorder-averaged free energy of the system by invoking the replica-symmetry ansatz.

\subsection{Replica trick for Gaussian-distributed random fields}
\label{sec:replica}

An established route to study quenched-disordered statistical systems is use of the so-called replica trick. The trick was first introduced by Anderson in 1975 in the context of spin-glass systems~\cite{Anderson_1975}, wherein the coupling between different spins is a random variable sampled from a Gaussian distribution. For such systems, this trick has proven to be immensely successful in characterizing the spin-glass phase. Here, we implement this trick for our class of models, by following the procedure for the mean-field version of the RFIM implemented by Schneider and Pytte in Ref.~\cite{Schneider-Pytte}. We consider the disordered fields to follow the Gaussian distribution~\eqref{eq:Gaussian}. 

\subsubsection{The replica trick}

By using the definition of the $\log$ function, one may write 
\begin{align}
{\label{eq:replica_trick_logarithm}}
\overline{\log \mathcal{Z} }= \lim_{p\to 0}\frac{\overline{\mathcal{Z}^p} - 1}{p}.
\end{align} 
In the replica trick, we consider several replicas of the same system. In each replica, the realization of quenched disorder $\{\Vec{h}_i\}$ is the same, but the dynamics of the spins may potentially lead to different equilibrium configurations, according to the multiple-equilibria picture proposed in the context of disordered systems~\cite{parisi_nobel_lecture}.

The replica trick was first proposed in the context of the random-bond Ising model with nearest-neighbor spin couplings that are random variables sampled from a Gaussian distribution. Random couplings lead to frustration in the system. Simply put, this refers to the effect that fixing the state of an initial spin and going around a loop while fixing the state of all spins one after another along the loop by respecting the sign of the spin-spin coupling, one may be required to return to
the initial spin and flip its state. As a result, at low temperatures and for a given disorder realization, the system possesses many configurations/states with similar energy that are separated by configurations with substantially higher energy. Consequently, prepared in one such low-energy state, low-temperature dynamics does not allow the system to access other close by states with similar energy owing to intervening large potential barriers. This leads to states within each local minima equilibrating locally with respect to a small set of energetically-accessible states. Systems prepared in each such local minima and equilibrating locally give rise to the concept of replicas that are identical copies of the same system with identical disorder realization but with potentially different equilibrium states. These states coincide when replica symmetry holds, but are different when there is spontaneous breaking of replica symmetry, as was first pointed out by Parisi for the spin-glass phase of the Sherrington-Kirkpatrick model~\cite{SK}. We now employ the replica trick to compute the disorder average of the product of the partition function, namely, the quantity $\overline{\mathcal{Z}^p}$.

\subsubsection{Computing the disorder average of the product of the partition function}
Using Eq.~\eqref{eq:replica_trick_logarithm}, we consider the product of the partition function for $p$ (integer) number of replicas and then take its disorder average, i.e., obtain the quantity $\overline{\mathcal{Z}^p}$, and finally, analytically continue the result to obtain the right hand side of Eq.~\eqref{eq:replica_trick_logarithm}. We have by definition that 
\begin{align}
{\label{eq12}}
\overline{\mathcal{Z}^p}&= \int \prod_{i=1}^N d\Vec{h}_i P(\Vec{h}_i) \int \prod_{a=1}^p D[\Vec{S}^{(a)}]  \nonumber \\
 &~~~~\times \exp\left[\frac{\beta J}{2N} \sum_{a=1}^p \sum_{i,j=1}^N \Vec{S_i}^{(a)} \cdot \Vec{S_j}^{(a)} + \beta \sum_{a=1}^p \sum_{i=1}^N \Vec{h_i} \cdot \Vec{S_i}^{(a)}\right],      \end{align}
where, as mentioned earlier, the index $a$ stands for the replicas, while the quantity $D[\Vec{S}^{(a)}]$ contains the relevant measure for $N$ spins in the $a$-th replica. 

Now, let us interchange the order of the two integrations in Eq.~\eqref{eq12}. Separating out the contribution due to a single spin-component, say, the $\alpha$-th component, of any spin in all the replicas, and integrating with respect to the measure of the same component of the random field acting on the same spin, we get 
\begin{align}
{\label{eq13}}
    &\int_{-\infty}^{\infty} \frac{dh_i^{(\alpha)}}{\sqrt{2 \pi \sigma^2}} \exp\left[\beta \left( \sum_{a=1}^p (S_i^{(\alpha)})^{(a)}\right) h_i^{(\alpha)} - \frac{{h_i^{(\alpha)}}^2}{2 \sigma^2}\right] \nonumber  \\
    &= \exp\left[\left(\sum_{a=1}^p (S_i^{(\alpha)})^{(a)}\right)^2 \frac{\beta^2\sigma^2}{2}\right]. 
\end{align}
Next, we use the Hubbard-Stratonovich identity
\begin{align} 
{\label{eq:Hubbard_Stratonovich}}
    \exp{(\lambda b^2)}= \int_{-\infty}^\infty \frac{dz}{\sqrt{2\pi}} \exp{\left(-\frac{z^2}{2} + \sqrt{2 \lambda} bz\right)},
\end{align}
with $\lambda$ real and positive, to write 
\begin{align}
 &\exp\left[\frac{\beta J}{2 N}\sum_{a=1}^p  \sum_{i,j=1}^N \Vec{S_i}^{(a)} \cdot \Vec{S_j}^{(a)}\right]\nonumber \\
 &= \int_{-\infty}^\infty \prod_{a=1}^p \frac{dz_a^{(\alpha)}}{\sqrt{2\pi}}~\exp\left[-\frac{1}{2} \sum_{a=1}^p (z_a^{(\alpha)})^2 \right] \nonumber \\
 &~~~~\times \exp\left[ \sqrt{\frac{\beta J}{N}} \sum_{a=1}^p\left(\sum_{i=1}^N (S_i^{(\alpha)})^{(a)}\right)z_a^{(\alpha)} \right],
 \label{eq:step1}
\end{align}
where $z_a^{(\alpha)}$'s are the independent auxiliary variables corresponding to $\sum_{i=1}^N (S_i^{(\alpha)})^{(a)}$. Note that obtaining Eq.~\eqref{eq:step1} relies on the all-to-all coupling between the spins. Using these results, and suppressing for brevity the explicit appearance of the integration range for each of the auxiliary variables $z_a^{(\alpha)}$, Eq.~\eqref{eq12} gives
\begin{align}
&\overline{\mathcal{Z}^p}\nonumber \\   
&=\int \prod_{\alpha=1}^n\prod_{a=1}^p \frac{dz_a^{(\alpha)}}{\sqrt{2\pi}}\exp\left[-\frac{1}{2} \sum_{\alpha=1}^n \sum_{a=1}^p (z_a^{(\alpha)})^2\right]\int \prod_{a=1}^p D[\Vec{S}^{(a)}]\nonumber \\
&~~~~\times \exp\Big[\sqrt{\frac{\beta J}{N}} \sum_{\alpha=1}^n \sum_{a=1}^p\left(\sum_{i=1}^N (S_i^{(\alpha)})^{(a)}\right)z_a^{(\alpha)}\nonumber \\
&~~~~+\sum_{\alpha=1}^n\sum_{i=1}^N \left(\sum_{a=1}^p (S_i^{(\alpha)})^{(a)}\right)^2 \frac{\beta^2\sigma^2}{2}\Big] \nonumber \\
&=\int \prod_{\alpha=1}^n\prod_{a=1}^p \frac{dz_a^{(\alpha)}}{\sqrt{2\pi}}\exp\left[-\frac{1}{2} \sum_{\alpha=1}^n \sum_{a=1}^p (z_a^{(\alpha)})^2\right]\left(\mathcal{K}[\{z_a^{(\alpha)}\}]\right)^N; \label{eq:auxiliary_partition_function}\\
&\mathcal{K}[\{z_a^{(\alpha)}\}]\equiv \int \prod_{a=1}^p D[\Vec{S}^{(a)}]\exp\Big[\sqrt{\frac{\beta J}{N}}\sum_{\alpha=1}^n \sum_{a=1}^p (S^{(\alpha)})^{(a)}z_a^{(\alpha)}\nonumber \\
&~~~~~~~~~~~~~~~~+\sum_{\alpha=1}^n\left(\sum_{a=1}^p (S^{(\alpha)})^{(a)}\right)^2 \frac{\beta^2\sigma^2}{2}\Big], \label{eq:single_spin_replica_partition} 
\end{align}    
where in the last equation, the quantity $D[\Vec{S}^{(a)}]$ now contains the relevant measure for a single spin in the $a$-th replica, while the quantity $\mathcal{K}[\{z_a^{(\alpha)}\}]$ may be interpreted as the single-spin partition function. 
In arriving at Eq.~\eqref{eq:auxiliary_partition_function} from Eq.~\eqref{eq12}, we have done a rearrangement of terms to have a trade-off between spin-spin interaction terms such as $(S_i^{(\alpha)})^{(a)}(S_j^{(\alpha)})^{(a)}$ and replica-replica interaction terms such as $(S_i^{(\alpha)})^{(a)}(S_i^{(\alpha)})^{(b)}$.

We now rescale the auxiliary variables as $z_a^{(\alpha)} \to (z_a^{(\alpha)})' = z_a^{(\alpha)}/\sqrt{N}$ and drop the primes, thereby obtaining
\begin{align} 
{\label{eq:replica_partition_saddle}}
\overline{\mathcal{Z}^p}&=  \int \prod_{\alpha=1}^n \prod_{a=1}^p \frac{dz_a^{(\alpha)} \sqrt{N}}{\sqrt{2\pi}}\nonumber\\
&~~~~\times \exp\left[-N\left(\frac{1}{2}\sum_{\alpha=1}^n \sum_{a=1}^p (z_a^{(\alpha)})^2  +  \log(\mathcal{K}[\{z_a^{(\alpha)}\}])\right)\right],
\end{align}
where we now have
\begin{align}
&\mathcal{K}[\{z_a^{(\alpha)}\}]\equiv \int \prod_{a=1}^p D[\Vec{S}^{(a)}]\exp\Big[\sqrt{\beta J}\sum_{\alpha=1}^n \sum_{a=1}^p (S^{(\alpha)})^{(a)}z_a^{(\alpha)}\nonumber \\
&~~~~~~~~~~~~~~~~+\sum_{\alpha=1}^n\left(\sum_{a=1}^p (S^{(\alpha)})^{(a)}\right)^2 \frac{\beta^2\sigma^2}{2}\Big]. \label{eq:single_spin_replica_partition-0} 
\end{align}

Let us invoke the saddle-point approximation for evaluation of the integral in Eq.~\eqref{eq:replica_partition_saddle}. We denote for the $a$-th replica and the $\alpha$-th spin component the minimizing auxiliary variable that minimizes the quantity multiplying the factor $N$ in the exponential in the above equation by $(z_a^{(\alpha)})_{\mathrm{min}}$, which evidently satisfies the following relation:
\begin{widetext}
\begin{align}\label{eqHreplica}
        (z_a^{(\alpha)})_{\mathrm{min}} & =  \sqrt{\beta J}\frac{\int \prod_{a=1}^p D[\Vec{S}^{(a)}] ~ (S^{(\alpha)})^{(a)} ~ \exp\left[\sqrt{\beta J} \sum_{\alpha=1}^n \sum_{a=1}^p(S^{(\alpha)})^{(a)} z_a^{(\alpha)}+\sum_{\alpha=1}^n\left(\sum_{a=1}^p (S^{(\alpha)})^{(a)}\right)^2 \frac{\beta^2\sigma^2}{2} \right]}{\int \prod_{a=1}^p D[\Vec{S}^{(a)}] ~ \exp\left[\sqrt{\beta J} \sum_{\alpha=1}^n \sum_{a=1}^p(S^{(\alpha)})^{(a)} z_a^{(\alpha)}+\sum_{\alpha=1}^n\left(\sum_{a=1}^p (S^{(\alpha)})^{(a)}\right)^2 \frac{\beta^2\sigma^2}{2} \right]}.
\end{align}
Under the scaling $z_a^{(\alpha)} \to  (z_a^{(\alpha)})' = z_a^{(\alpha)}/\sqrt{\beta J}$, and dropping the primes, we get
\begin{align}\label{eqHreplica-1}
        (z_a^{(\alpha)})_{\mathrm{min}} & =  \frac{\int \prod_{a=1}^p D[\Vec{S}^{(a)}] ~ (S^{(\alpha)})^{(a)} ~ \exp\left[\beta J \sum_{\alpha=1}^n \sum_{a=1}^p(S^{(\alpha)})^{(a)} z_a^{(\alpha)}+\sum_{\alpha=1}^n\left(\sum_{a=1}^p (S^{(\alpha)})^{(a)}\right)^2 \frac{\beta^2\sigma^2}{2} \right]}{\int \prod_{a=1}^p D[\Vec{S}^{(a)}] ~ \exp\left[\beta J \sum_{\alpha=1}^n \sum_{a=1}^p(S^{(\alpha)})^{(a)} z_a^{(\alpha)}+\sum_{\alpha=1}^n\left(\sum_{a=1}^p (S^{(\alpha)})^{(a)}\right)^2 \frac{\beta^2\sigma^2}{2} \right]}.
\end{align}
\end{widetext}

\begin{figure}
    \centering
    \includegraphics[width=1\linewidth]{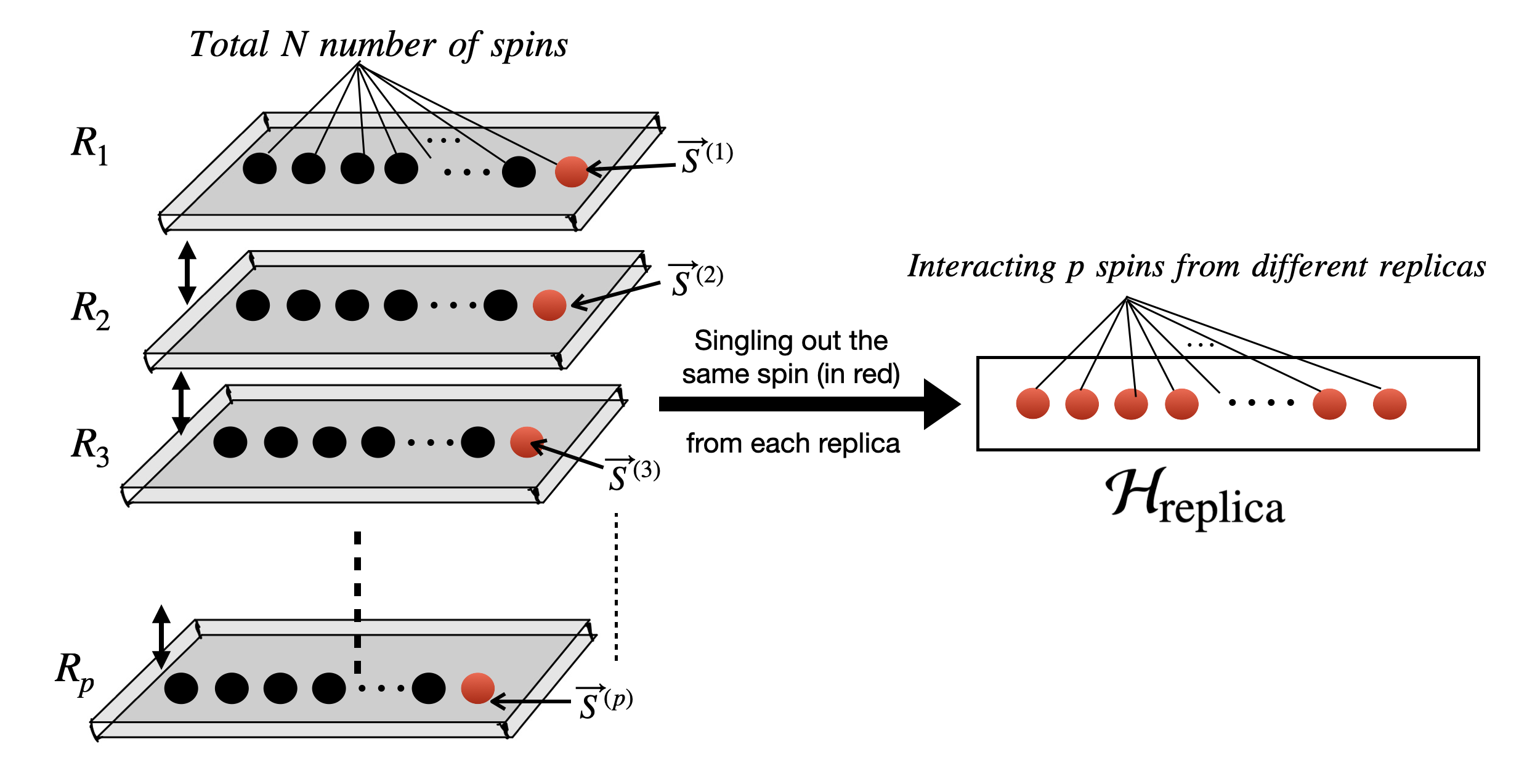}
    \caption{Constructing the system~\eqref{replica-system}: For the system~\eqref{eq:H} with the disordered fields sampled from a Gaussian distribution, we first consider $p$ replicas of the system, namely, $R_1$, $R_2$, $\ldots$, $R_p$, in each of which the disorder realization is the same. We single out the same spin (in red) from each replica, and construct a new physical system comprising these spins $\{\Vec{S}^{(a)}\}_{1\le a \le p}$ with the Hamiltonian $\mathcal{H}_{\mathrm{replica}}$ given in Eq.~\eqref{replica-system}. In this new system, each spin is subjected to a different onsite field $J\vec{z}_{a}$.} 
    \label{fig:replica-system-construction}
\end{figure}

In order to understand physically the right hand side of Eq.~\eqref{eqHreplica-1}, we now move on to construct a Hamiltonian in the replica space, namely, the replica Hamiltonian.

\subsubsection{Constructing the replica Hamiltonian}
The physical interpretation of the right hand side of Eq.~\eqref{eqHreplica-1} is the following. Let us single out one spin (the same spin) from each of the $p$ replicas, and construct a new physical system comprising these spins $\{\Vec{S}^{(a)}\}_{1\le a \le p}$, with each subjected to an onsite field $J\vec{z}_{a}$. Let us define the Hamiltonian of the system as (see Fig.~\ref{fig:replica-system-construction}) 
\begin{align}\label{replica-system}
    \mathcal{H}_\mathrm{replica}=J  \sum_{a=1}^p \Vec{S}^{(a)} \cdot \vec{z}_a+\frac{\beta\sigma^2}{2}\sum_{a,b=1}^p \Vec{S}^{(a)}\cdot \Vec{S}^{(b)}. 
\end{align}
For $\sigma=0$, computing the canonical-equilibrium average with respect to $\mathcal{H}_\mathrm{replica}$, we obtain  
\begin{align}
        &\langle (S^{(\alpha))})^a\rangle_{\mathcal{H}_\mathrm{replica}}\nonumber \\
        &=\frac{\int \prod_{a=1}^p D[\Vec{S}^{(a)}] ~ (S^{(\alpha)})^{(a)} ~ \exp\left[\beta \mathcal{H}_\mathrm{replica}(\sigma=0)\right]}{\int \prod_{a=1}^p D[\Vec{S}^{(a)}]~\exp\left[\beta \mathcal{H}_\mathrm{replica}(\sigma=0)\right]}\nonumber \\
        &=\langle S^{(\alpha)}\rangle_{\mathcal{H}_\mathrm{replica}}=m^{(\alpha)},
\end{align}
which coincides with the right hand side of Eq.~\eqref{eqHreplica-1} with $\sigma=0$. In arriving at the last equality, we have used the fact that for $\sigma=0$, the different spins in our system, coming from different replicas, are not interacting with one another. Here, we have denoted by $m^{(\alpha)}$ the thermal average of the $\alpha$-th spin component with respect to the canonical-equilibrium measure with respect to the Hamiltonian $\mathcal{H}_\mathrm{replica}$ and at temperature $1/\beta$. For $\sigma\ne 0$, we may then write on using Eq.~\eqref{eqHreplica-1} that 
\begin{align}
        (z_a^{(\alpha)})_{\mathrm{min}} & = {\langle(S^{(\alpha)})^{(a)}\rangle}_{\mathcal{H}_\mathrm{replica}}=(m^{(\alpha)})^{(a)}.
\end{align}

In the following, we proceed with our analysis by invoking the replica-symmetry ansatz.

\subsubsection{The replica-symmetry ansatz: Obtaining the disorder-averaged free energy}
In principle, the values of the minimizing auxiliary variables $(z_a^{(\alpha)})_{\mathrm{min}}$ could be different for different replicas. However, let us invoke the replica-symmetry ansatz, which demands that these independent minimizing auxiliary variables take up the same value across the different replicas, i.e., 
\begin{align}
(z_a^{(\alpha)})_{\mathrm{min}} =(z_b^{(\alpha)})_{\mathrm{min}} ;~a\neq b;~a,b \in \{1,\ldots,p\}. 
\label{eq:ansatz-z}
\end{align}
Physically, the ansatz implies that the fluctuations of ${\langle(S^{(\alpha)})^{(a)}\rangle}_{\mathcal{H}_\mathrm{replica}}$ across the different spins in the system~\eqref{replica-system} are negligible, and one has 
\begin{align}{\label{eq:RSA-1}}
    (z_1^{(\alpha)})_{\mathrm{min}}=(z_2^{(\alpha)})_{\mathrm{min}}= \ldots=(z_p^{(\alpha)})_{\mathrm{min}} =m^{(\alpha)},
\end{align}
which in terms of unscaled $z_a^{(\alpha)}$ (see the discussion following Eq.~\eqref{eqHreplica}) reads as 
\begin{align}{\label{eq:RSA}}
    (z_1^{(\alpha)})_{\mathrm{min}}=(z_2^{(\alpha)})_{\mathrm{min}}= \ldots=(z_p^{(\alpha)})_{\mathrm{min}} =   \sqrt{\beta J} m^{(\alpha)}.
\end{align}
One may view the ansatz in Eq.~\eqref{eq:ansatz-z} from another perspective. Had the fields $\Vec{h}_i$ in Eq.~\eqref{eq:H} been the same for all spins (i.e., had there been no disorder in the system), all the different replicas would have identical equilibrium configurations. The ansatz asserts that the same remains true even when there is disorder in the system.   

Putting back Eq.~\eqref{eq:RSA} in Eq.~\eqref{eq:single_spin_replica_partition-0} gives
\begin{align}{\label{eq:RSA-1}}
    &\mathcal{K}[\{(z_a^{(\alpha)})_{\mathrm{min}}\}]= \int \prod_{a=1}^p D[\Vec{S}^{(a)}] \nonumber \\
    &~~~~\times\exp\left\{ \sum_{\alpha=1}^n\left[ \beta Jm^{(\alpha)} \left(  A^{(\alpha)}\right) + \frac{\beta^2 \sigma^2}{2} \left(  A^{(\alpha)}\right)^2 \right]\right\},
\end{align}
where we have defined $A^{(\alpha)} \equiv   \sum_{a=1}^p (S^{(\alpha)})^{(a)}$. Now, we use the Hubbard-Stratonovich identity in Eq.~\eqref{eq:Hubbard_Stratonovich} to write 
\begin{align}
    &\exp{\left[\frac{\beta^2 \sigma^2}{2} \left(  A^{(\alpha)}\right)^2 \right]} \nonumber \\
    &= \int_{-\infty}^\infty \frac{dh^{(\alpha)}}{\sqrt{2\pi \sigma^2}} \exp{\left[ -\frac{(h^{(\alpha)})^2}{2\sigma^2} + \beta A^{(\alpha)} h^{(\alpha)}\right]}, 
\end{align}
which on using Eq.~\eqref{eq:RSA-1} results in
\begin{align}\label{eq:single_spin_replica_partition-1}
\mathcal{K}[\{(z_a^{(\alpha)})_{\mathrm{min}}\}]& = \int \prod_{a=1}^p D[\Vec{S}^{(a)}] \int  d\Vec{h}~P(\Vec{h})\nonumber \\
&~~~~\times \exp\left\{ \sum_{\alpha=1}^n\left[ \beta (Jm^{(\alpha)} + h^{(\alpha)}) A^{(\alpha)}  \right]\right\}\nonumber \\
    & = \int d\Vec{h}~P(\Vec{h}) \int \prod_{a=1}^p D[\Vec{S}^{(a)}]\nonumber \\
    &~~~~\times \exp\left\{ \sum_{\alpha=1}^n\left[ \beta (Jm^{(\alpha)} + h^{(\alpha)}) A^{(\alpha)}  \right]\right\};
\end{align}
$\int P(\Vec{h}) d\Vec{h}= \int \prod_{\alpha=1}^n dh^{(\alpha)}
/\sqrt{2\pi\sigma^2}\exp{\left( - (h^{(\alpha)})^2/(2\sigma^2)\right)}$. Also, note that the order of integrals has been interchanged in order to arrive at the last equality. After separating the different replica contributions in the exponential, as in  
\begin{align}
    &\exp\left\{ \sum_{\alpha=1}^n \beta (Jm^{(\alpha)} + h^{(\alpha)}) A^{(\alpha)}  \right\} \nonumber \\
    &= \exp\left\{ \sum_{a=1}^p \sum_{\alpha=1}^n \beta (Jm^{(\alpha)} + h^{(\alpha)}) (S^{(\alpha)})^{(a)}  \right\}\nonumber \\
    & = \exp\left\{ \sum_{a=1}^p  \beta (J\Vec{m} + \Vec{h}) \cdot \Vec{S}^{(a)}  \right\},
\end{align}
we can write the joint integral on the left hand side of the following as a product of $p$ identical integrals:  
\begin{align}
    &\int \prod_{a=1}^p D[\Vec{S}^{(a)}] \exp\left\{ \sum_{\alpha=1}^n \beta (Jm^{(\alpha)} + h^{(\alpha)}) A^{(\alpha)}  \right\}\nonumber \\
    &= \left(\int d\Vec{S} \exp[\beta (J\Vec{m} + \Vec{h})\cdot\Vec{S}] \right)^p,
\end{align}
since the different $\Vec{S}^{(a)}$'s are independent and identical. 
Using the above expression, we finally arrive from Eq.~\eqref{eq:single_spin_replica_partition-1} to 
\begin{align}
\mathcal{K}[\{(z_a^{(\alpha)})_\mathrm{min}\}]= \int d\Vec{h}
~P(\Vec{h}) \left(\int d\Vec{S} \exp[\beta (\Vec{h}+J\Vec{m})\cdot\Vec{S} \right)^p.
\end{align}
This completes the computation of $\overline{\mathcal{Z}^p}$, for which, using Eq.~\eqref{eq:replica_partition_saddle} and the saddle-point approximation, we now have the final expression: 
\begin{align}
    \overline{\mathcal{Z}^p} = \exp\left[-\frac{N\beta J p}{2}  \sum_{\alpha=1}^n(m^{(\alpha)})^2  + N \log(\mathcal{K}_1[\{z_a^{(\alpha)}\}_\mathrm{min}])\right]. 
\end{align}

As suggested by Eq.~\eqref{eq:replica_trick_logarithm}, we express the free energy $\mathcal{F}=-(1/\beta)\, \overline{\log \mathcal{Z} }$ as 
   \begin{align}{\label{eq:Free_energy_replica_limit}}
    -\beta \mathcal{F} &= \lim_{p \to 0} \frac{1}{p}~\Big[\exp\Big(N \Big\{-\frac{\beta Jp}{2} \sum_{\alpha=1}^n (m^{(\alpha)})^2 \nonumber \\
    &~~~~+ \log\left[\mathcal{K}\left( \{ (z_a^{(\alpha)})_\mathrm{min} \}\right)\right]   \Big\}\Big)-1 \Big].
\end{align}
After implementing the limit in Eq.~\eqref{eq:Free_energy_replica_limit} using the L'H\^{o}spital's rule, we get the free energy per spin as 
\begin{align}{\label{eq:replica_symmetric_free_energy}}
\frac{\mathcal{F}}{N}&=\frac{Jm^2}{2} -   \frac{1}{\beta}\int d{\Vec{h}}~ P(\Vec{h})   \log\left\{\int d \Vec{S} \exp[\beta \Vec{S} \cdot (\Vec{h} + J\Vec{m})]   \right\}. \end{align}

\subsubsection{Comparing disorder-averaged free energy expressions obtained from mean-field approximation and replica trick}

Clearly, Eq.~\eqref{eq:replica_symmetric_free_energy} for the replica-symmetric expression of free energy per spin, for the case of disordered fields sampled from the Gaussian distribution, coincides with the one obtained from the  mean-field approximation, Eq.~\eqref{eq_F}. However, one should note that the replica trick could be employed only for the Gaussian distribution, but the expression in Eq.~\eqref{eq_F} is expected to hold true for any distribution without broad tails.

\begin{figure}
    \centering
    \includegraphics[width=1.\linewidth]{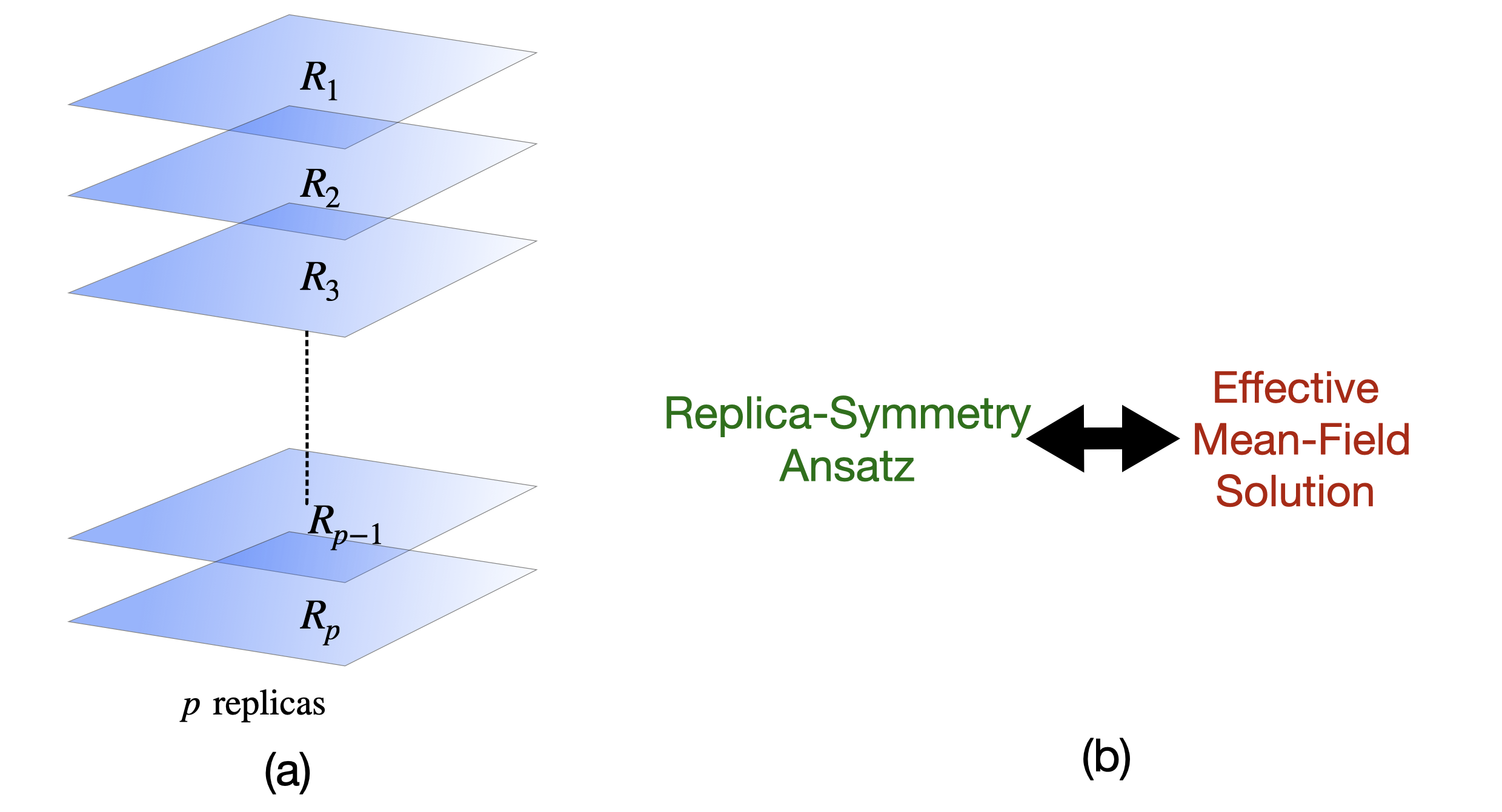}
    \caption{Panel~(a) demonstrates a visual representation of the replica trick in Eq.~\eqref{eq:replica_trick_logarithm}, wherein the blue sheets labeled $\{R_1, R_2, \ldots, R_p\}$  are the $p$ replicas. Panel~(b) represents a schematic of the outcome of our analysis, showing that when the disorder distribution is Gaussian, the replica-symmetric-ansatz solution for the free energy matches the one obtained from a mean-field approximation.}
    \label{fig:replica_symmetry}
\end{figure}

The aforementioned matching of free-energy expressions suggests that replica-symmetry breaking, which is a key feature in spin-glass systems, is absent for our class of mean-field $O(n)$ models with disordered fields, at least when the fields are sampled from a Gaussian distribution. The stability of the replica-symmetry ansatz for $n=2$ has also been discussed in Ref.~\cite{Dries_XY}, and the solution is shown to be marginally stable. The key points of our analysis pursued thus far are pictorially summarized in Fig.~\ref{fig:replica_symmetry}.

\section{Phase transitions in the $n=2$ case}
\label{sec:n2}
Now, we discuss the effect of disorder on the critical point of mean-field $O(n)$ systems. In this section, we consider the case of $n=2$, which is the classical $XY$ spin model, and which serves as a representative model to carry out further analytical studies. The choice of distributions for disordered fields will be the Gaussian and bimodal distributions defined in Section~\ref{sec:model}.

The spin components for the $n=2$ case are $(S^x, S^y) = (\cos{\theta}, \sin{\theta})$, with $0 \leq\theta<2\pi $. The free energy per spin is obtained from Eq.~\eqref{eq_F} as 
\begin{align}
{\label{eq:XY_free_energy}}
\frac{\mathcal{F}}{N} &= \frac{J(m_x^2 + m_y^2)}{2} \nonumber \\
&~~~~- \frac{1}{\beta} \int d\Vec{h}~P(\Vec{h})  \log\left[\int_{0}^{2\pi} d\theta~ \exp\left(\beta \Vec{S} \cdot (\Vec{h} + J\Vec{m})\right)\right],
\end{align}
with $\Vec{m}=(m_x, m_y)$. Using the identity 
\begin{align}{\label{eq:Bessel_zeroth_idenity}}
    \frac{1}{2\pi}\int_{0}^{2\pi} d\theta \exp{\left( a \cos{\theta} + b\sin{\theta}\right)} = I_0 \left(\sqrt{a^2 + b^2}\right),
\end{align}
where $I_0(x)$ is the zeroth-order modified Bessel function of the first kind, we can write from Eq.~\eqref{eq:XY_free_energy} that 
\begin{align}{\label{eq:XY_modified_free_energy}}
\frac{\mathcal{F}}{N} &= \frac{J(m_x^2 + m_y^2)}{2}\nonumber \\
&~~~~-\frac{1}{\beta} \int d\Vec{h}~P(\Vec{h})  \log\left[2\pi I_0\left( \beta\sqrt{\sum_{\alpha=x,y} (h^{(\alpha)} + J m^{(\alpha)})^2}\right)\right]. 
\end{align}\\
We now proceed to consider the two mentioned disorder distributions. 
\begin{figure}
    \centering
    \includegraphics[width=1.0\linewidth]{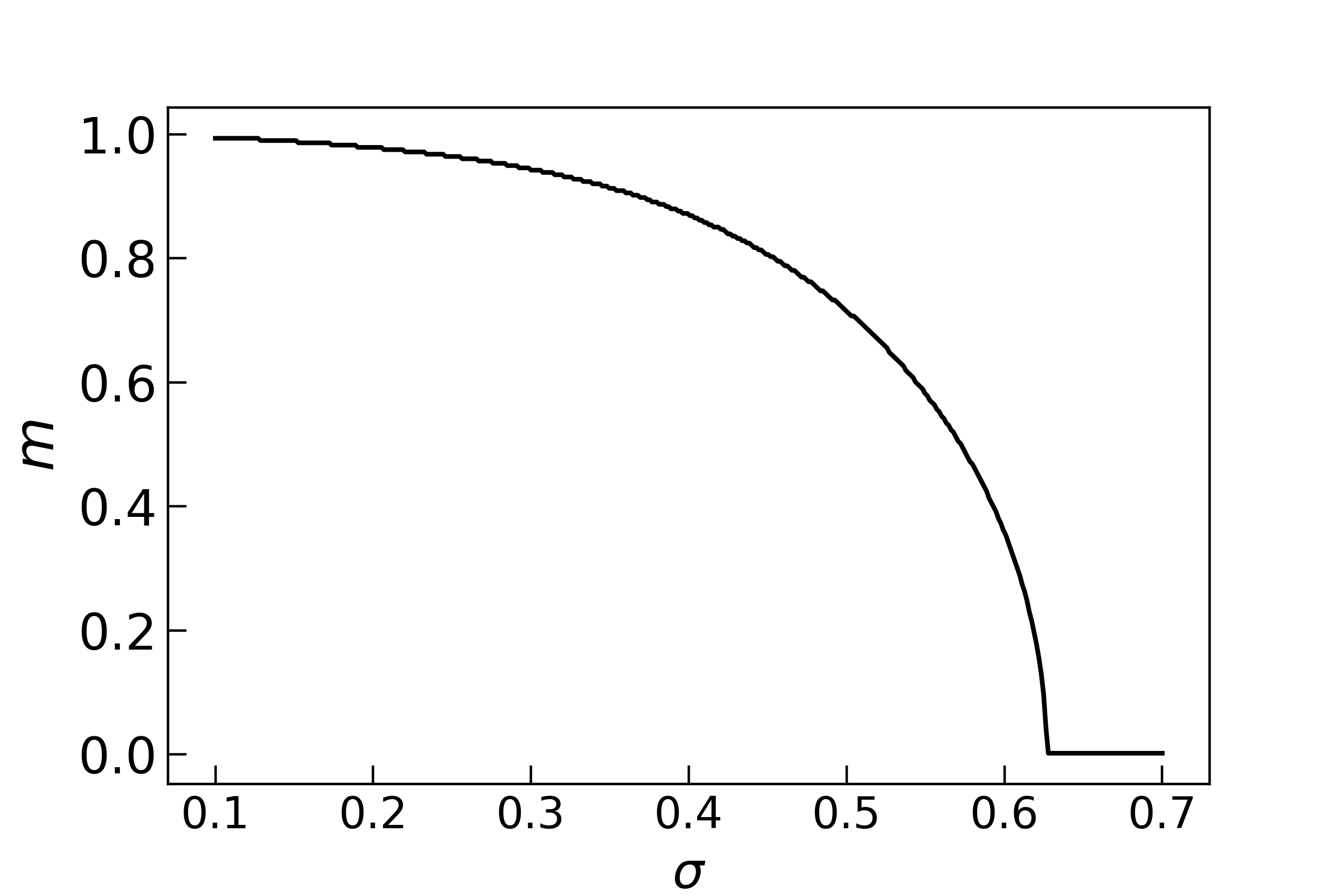}
    \caption{For the case $n=2$ in Eq.~\eqref{eq:H}, corresponding to the $XY$ spin model with a Gaussian disorder distribution, the system at $T=0$ undergoes a continuous phase transition in equilibrium magnetization $m$ on tuning the disorder strength $\sigma$. We consider here $J=1$. The critical disorder strength, identified as the value of $\sigma$ at and above which the magnetization value is zero, is given by $\sigma_c \approx 0.6266$; a detailed discussion on numerical estimation of this value is given in Appendix~\ref{apperror}. The data in this figure are generated by numerically extracting the global minimum of Eq.~\eqref{eq:Zero_temp_XY_free_energy_Gaussian}.}
    \label{fig:XY_Gaussian_ground_state_m_vs_sigma}
\end{figure}

\begin{figure}
    \centering
    \includegraphics[width=1\linewidth]{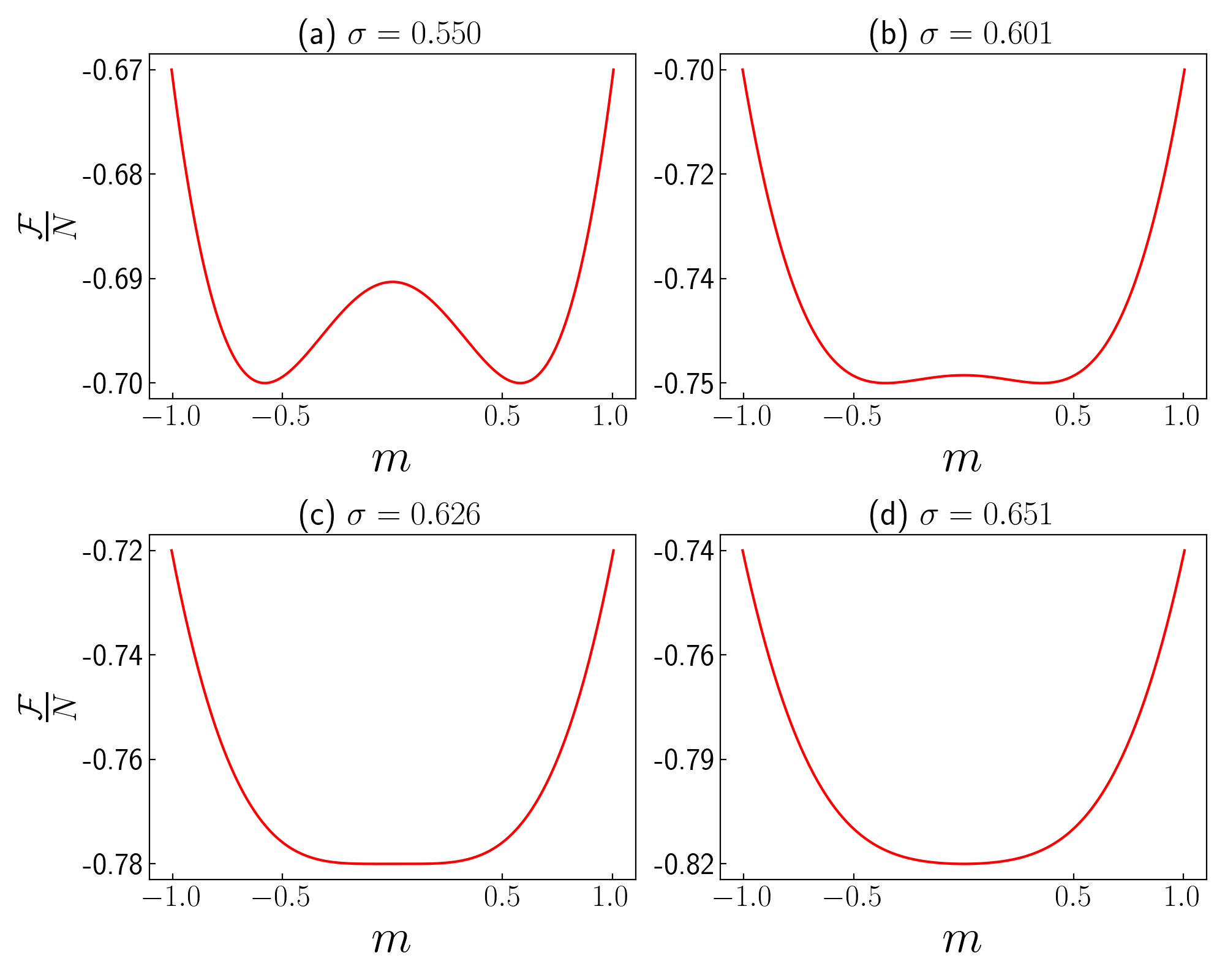}
    \caption{Ground state free energy landscape, Eq.~\eqref{eq:Zero_temp_XY_free_energy_Gaussian}, in the vicinity of the critical point $\sigma_c = 0.6266J$. We take $J =1$.}
    \label{fig:fe}
\end{figure}
\subsubsection{Gaussian distribution}
Using Eq.~\eqref{eq:Gaussian} for the case at hand, the Gaussian distribution for the disordered field $\Vec{h}_i$ is 
\begin{align}
P(\Vec{h}) &  = \frac{1}{2 \pi \sigma^2} \exp{\left(- 
 \frac{{(h^{(x)}})^2 + {(h^{(y)}})^2}{2 \sigma^2}\right) } \nonumber \\
 & = \frac{1}{2 \pi \sigma^2}~ \exp{\left(- \frac{|\Vec{h}|^2}{2 \sigma^2}\right)},
\end{align}
with $-\infty \leq h^{(x)}, h^{(y)} \leq \infty$. Now remembering that the magnetization order parameter is $m = \sqrt{m_x^2 + m_y^2}$, we use the transformation $v_x \equiv h^x + J m_x, v_y \equiv h^y + J m^y$, to write $\sum_{\alpha=x,y} (h^{(\alpha)} + J m^{(\alpha)})^2 = v_x^2 + v_y^2$, and that 
\begin{align}
&(h^{(x)})^2 + (h^{(y)})^2 \nonumber \\
&= (v_x^2 + v_y^2) + J^2(m_x^2 + m_y^2) - 2J(v_x m_x + v_y m_y).
\end{align}
Let us resort to polar coordinates for $\Vec{v}$, and write $(v_x,v_y)= (r \cos{\phi},r\sin{\phi})$, with $r \in [0,\infty)$ and $0 \le \phi <2\pi$. This allows to write Eq.~\eqref{eq:XY_modified_free_energy} as 
\begin{align}
    \frac{\mathcal{F}}{N} &= \frac{Jm^2}{2} - \frac{\exp{\left(-\frac{J^2 m^2}{2\sigma^2}\right)}}{\beta} \int_{0}^{\infty}dr~\frac{r}{2\pi \sigma^2} \exp{\left(-\frac{r^2}{2\sigma^2}\right)} \nonumber \\  
    &\quad \times \log{[2\pi I_0(\beta r)]}\int_{0}^{2\pi} d\phi~ \exp\left[\frac{Jr(m_x \cos{\phi} + m_y \sin{\phi})}{\sigma^2}\right].
\end{align}

\begin{figure}
    \centering
    \includegraphics[width=1.0\linewidth]{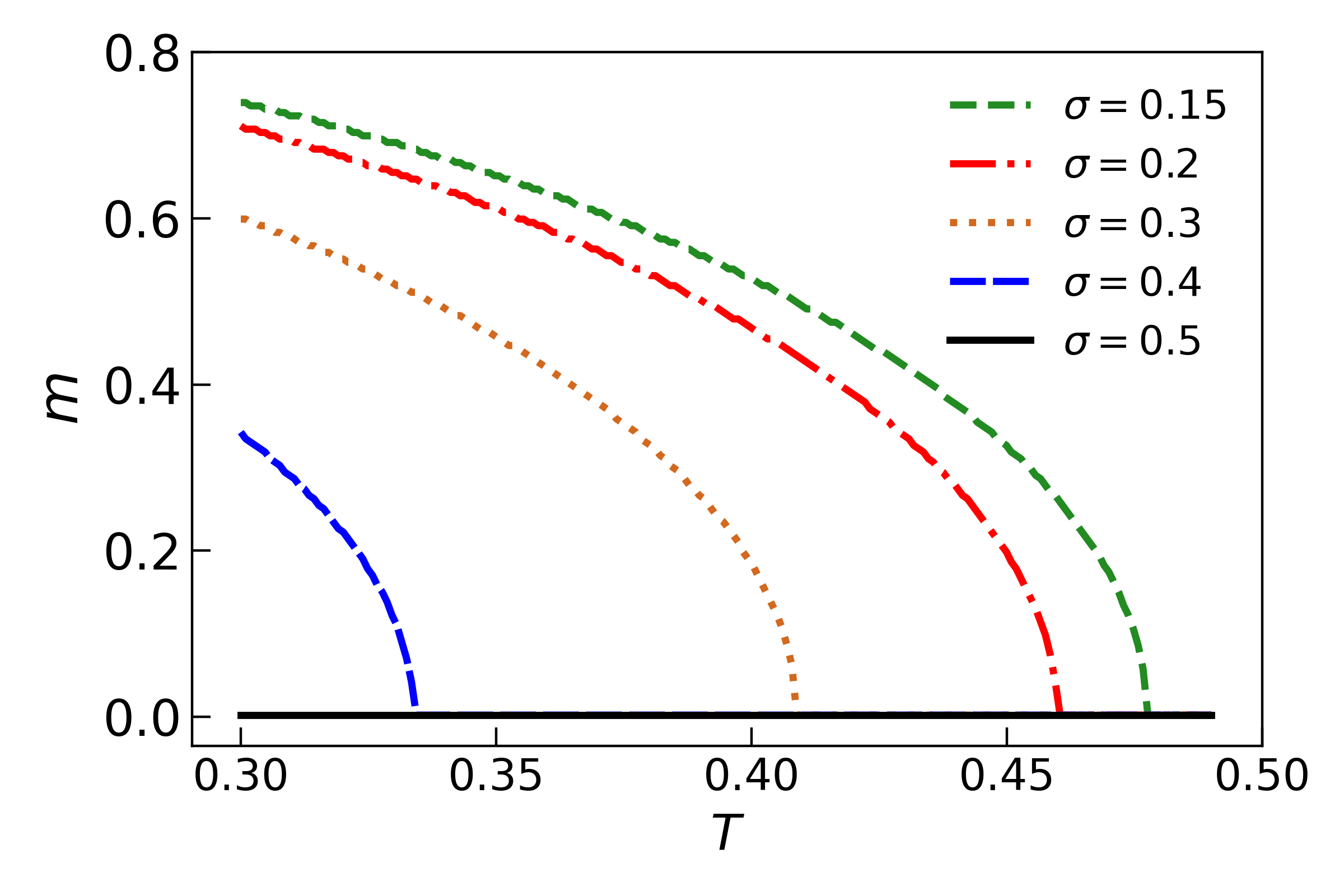}
    \caption{For the case $n=2$ in Eq.~\eqref{eq:H}
    , i.e., for $XY$ spins with a Gaussian disorder distribution and with $T>0$, the variation of equilibrium magnetization $m$ as a function of temperature is shown for various disorder strength $\sigma$. We consider here $J=1$. The figure clearly demonstrates that even with non-zero disorder, the equilibrium magnetization undergoes a continuous phase transition, similar to the case without disorder. However, as expected, the critical temperature decreases as the disorder strength increases. This behavior aligns with the general understanding that disorder tends to suppress order. The data in this figure are generated by numerically extracting the global minimum of Eq.~\eqref{eq:free_energy_XY_rotational}.}
    \label{fig:XY_Gaussian_m_vs_T_theory}
\end{figure}

\begin{figure*}
    \centering
    \includegraphics[width=\textwidth]{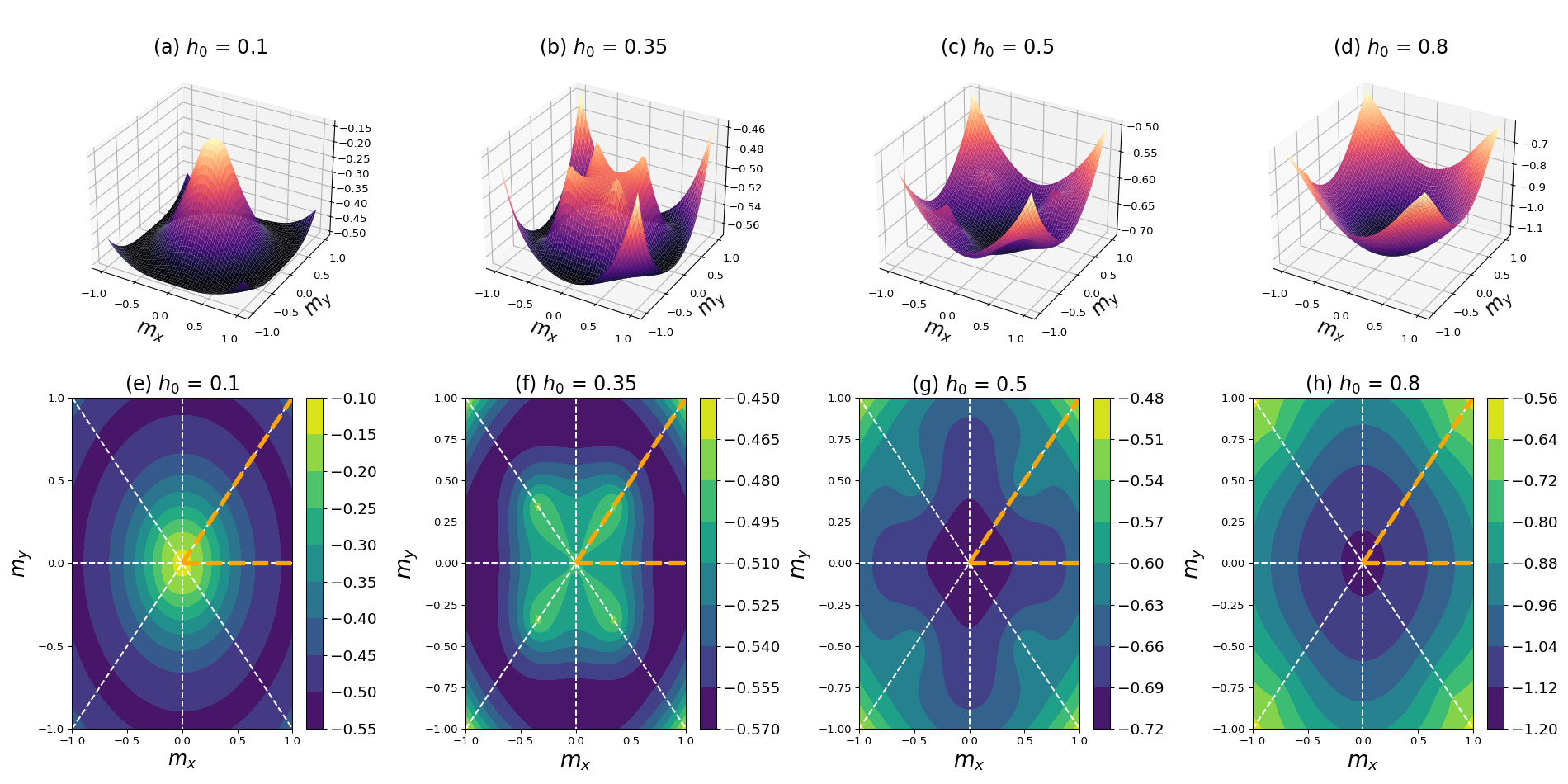}
    \caption{For the case $n=2$ in Eq.~\eqref{eq:H}, i.e., for the $XY$ spin model with bimodal disorder distribution as in Eq.~\eqref{eq:bimodal}, the zero-temperature free-energy landscape of Eq.~\eqref{eq:XY_bimodal_ground_state_free_energy} is shown in the $(m_x,m_y)$-plane for a few representative disorder strengths $h_0$. We consider here $J=1$. Clearly, for very small and large $h_0$, the landscape has a single minimum as in panels (a), (e), and panels (d), (h), respectively. At intermediate $h_0$ values, the free-energy landscape admits multiple minima, and with increasing disorder strength, the global minimum, identified with the equilibrium magnetization $m=\sqrt{m_x^2 + m_y^2}$, changes from non-zero to zero. The white dashed lines in panels (e) -- (h) divide the landscape into regions in which the free energy has identical behavior, while the orange lines bound the region in which we perform numerical minimization of the free energy.}
    \label{fig:XY_bimodal_ground_state}
\end{figure*}

Under the scaling $r \to r/\sigma $ and using the identity in Eq.~\eqref{eq:Bessel_zeroth_idenity}, one may express the disorder-averaged free energy per spin as 
\begin{align} {\label{eq:free_energy_XY_rotational}}
    &\frac{\mathcal{F}}{N}= \frac{Jm^2}{2}- \frac{1}{\beta} \exp{\left(-\frac{J^2 m^2}{2\sigma^2}\right)}\nonumber \\
    &\times \int_{0}^{\infty} dr~r \exp{\left(-\frac{r^2}{2}\right)}~\log{[2\pi I_0(\beta r \sigma)]}~I_0\left(\frac{Jmr}{\sigma}\right). 
\end{align}
\begin{figure}[b]
    \centering
    \includegraphics[width=1\linewidth]{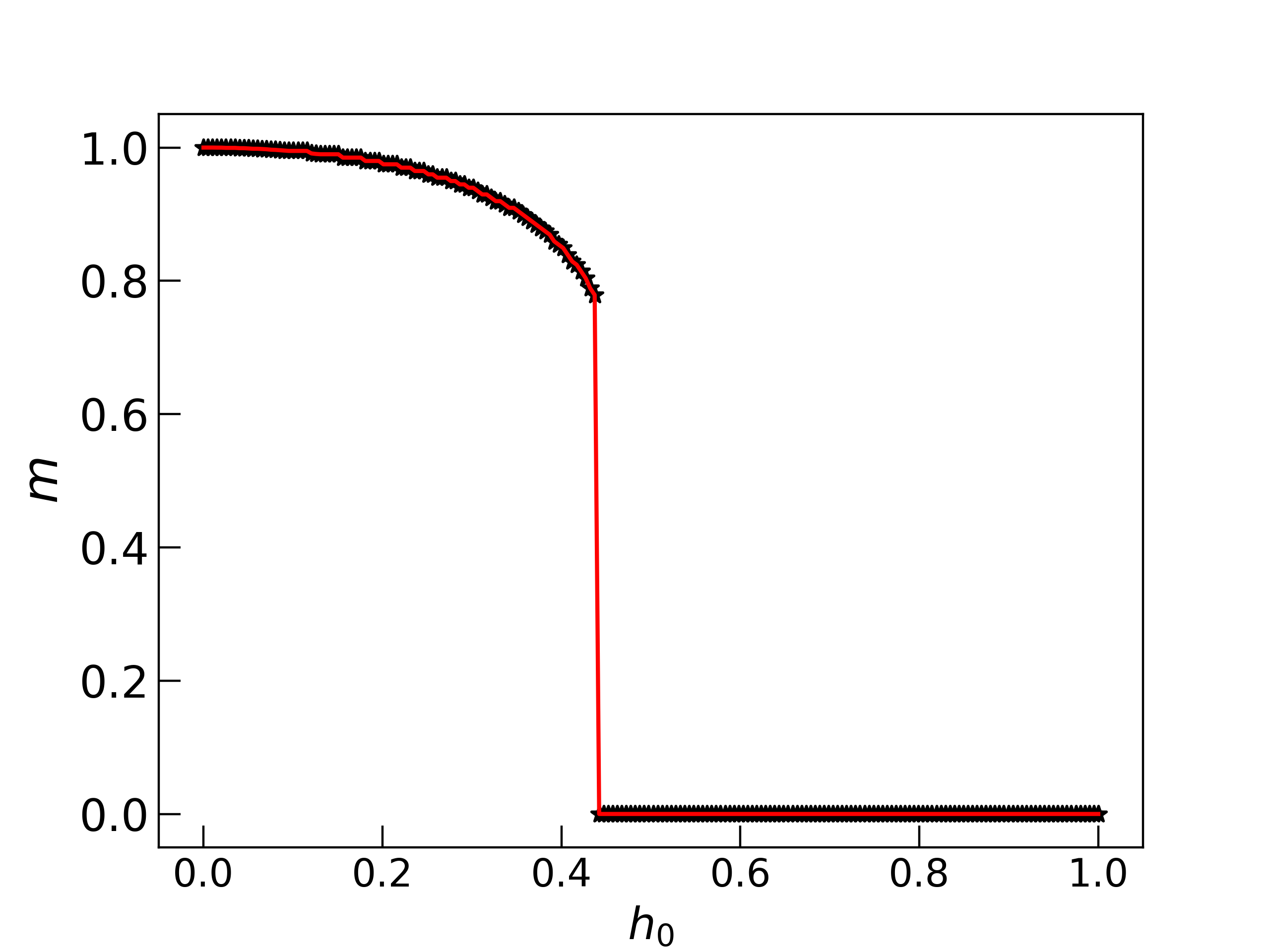}
    \caption{For the case $n=2$ in Eq.~\eqref{eq:H}, corresponding to the $XY$ spin model with bimodal disorder distribution, the $T=0$ equilibrium magnetization (i.e., the ground-state magnetization) $m$ of the system undergoes a first-order phase transition from a ferromagnetic to a paramagnetic phase with increasing disorder strength $h_0$, indicated by the sudden jump in magnetization $m$ at the critical value of disorder given by $h_0^c\approx 0.4422$ (we consider here $J=1$); a detailed discussion on numerical estimation of this value of $h_0^c$ is given in Appendix~\ref{apperror}. The data in this figure are generated by numerically extracting the global minimum of Eq.~\eqref{eq:XY_bimodal_ground_state_free_energy}.}
    \label{fig:XY_bimodal_ground_m_vs_h0}
\end{figure}

\begin{figure*}
    \centering
    \includegraphics[width=\linewidth]{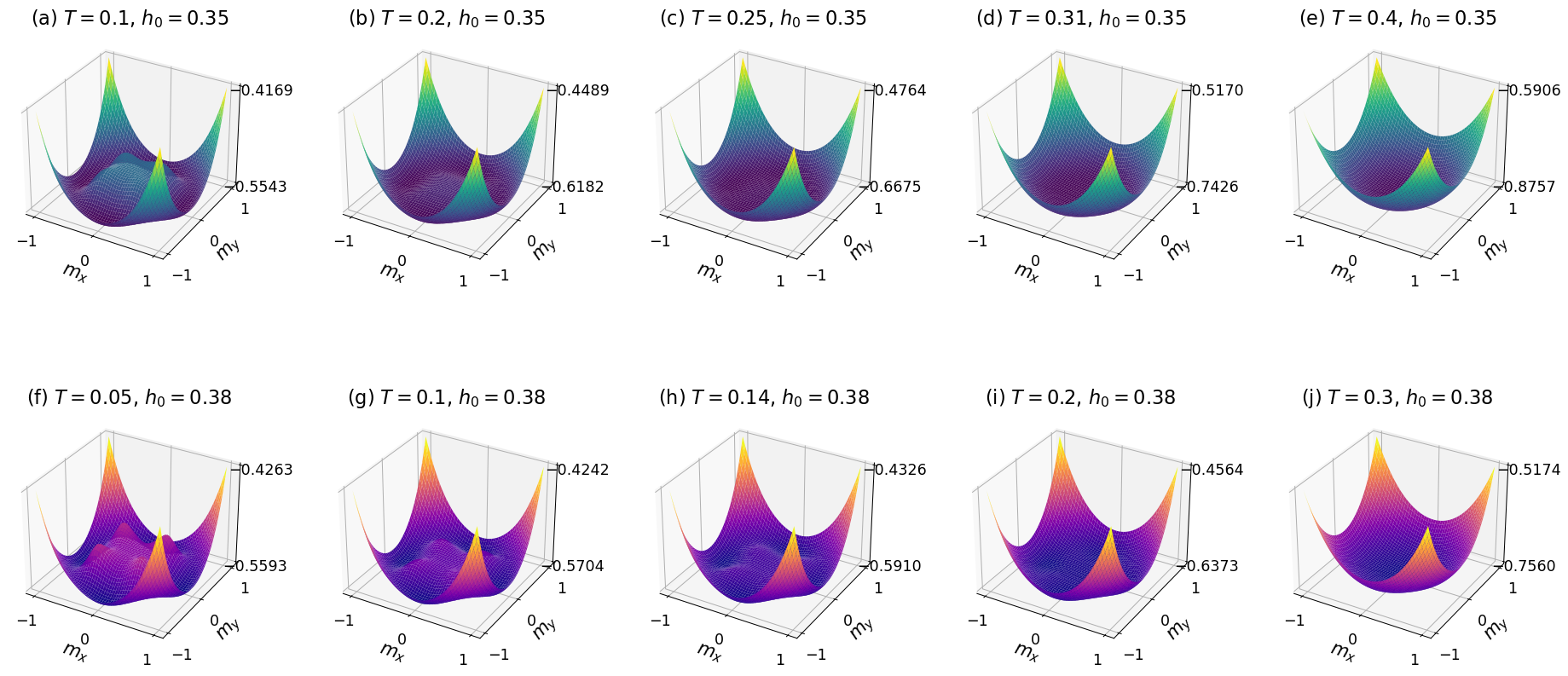}
    \caption{For the case $n=2$ in Eq.~\eqref{eq:H}, i.e., for the $XY$ spin model with bimodal disorder distribution as in Eq.~\eqref{eq:bimodal}, the finite temperature free-energy landscape of Eq.~\eqref{eq:XY_bimodal_free_energy} is shown in the $(m_x,m_y)$-plane for two representative disorder strengths $h_0$. We consider here $J=1$. The plots in panels (a)-(e) correspond to the disorder strength $h_0= 0.35$, while those in panels (f)-(j) correspond to the disorder strength $h_0=0.38$, with the temperature in each case increasing from left to right. Clearly, for panels in both top and bottom rows, the landscape has a single global minimum for small and large $T$,  while at intermediate $T$, the landscape admits multiple minima. The landscapes shown in the top panel correspond to the system approaching the tricritical point from the side of continuous transition, whereas those in the bottom panel depict the system deep in the first-order transition region.}
    \label{fig:XY_bimodal_finite_temp_free_energy_landscape}
\end{figure*}

Following the standard prescription of Landau theory, we proceed to expand the free-energy expression in Eq.~\eqref{eq:free_energy_XY_rotational} around $m=0$. Let us first note that for small $m$, one has~\cite{dlmf} 
\begin{align}\label{eq:Bessel function expansion around zero}
    I_0\left(\frac{Jmr}{\sigma}\right) = 1 + \frac{J^2 m^2 r^2}{4\sigma^2} + \frac{J^4 m^4 r^4}{64 \sigma^4} + \ldots. 
\end{align}
Using the above expansion in Eq.~\eqref{eq:free_energy_XY_rotational}, we get
\begin{align}
    &\frac{\mathcal{F}}{N}= \frac{Jm^2}{2} - \frac{1}{\beta} \int_{0}^{\infty} dr~r \exp{(-\frac{r^2}{2})}~\log{[2 \pi I_0(\beta r \sigma)]}\nonumber \\
    &\times \left[ 1 - \frac{J^2m^2}{2\sigma^2} + \frac{J^4m^4}{8\sigma^4} + \ldots \right] \left[ 1 + \frac{J^2 m^2 r^2}{4\sigma^2} + \frac{J^4 m^4 r^4}{64 \sigma^4} + \ldots \right] \nonumber \\
    & = - \frac{1}{\beta} [r] + m^2 \left( \frac{J}{2} - \frac{J^2}{2\beta\sigma^2} [r] + \frac{J^2}{4 \beta \sigma^2} [r^3]\right) \nonumber \\
    &+ m^4\left(- \frac{J^4}{64\beta \sigma^4} [r^5]  - \frac{J^4}{8\beta \sigma^4} [r]+ \frac{J^4}{8\beta \sigma^4} [r^3]   \right) + \ldots \nonumber\\
    & = a_0 (T) + a_2(T) m^2 +  a_4(T) m^4 + \ldots,
\end{align} 
where we have used the definition 
\begin{align}
    [r^k] \equiv \int_{0}^\infty dr ~r^k \exp{\left(-\frac{r^2}{2}\right)} \log{\left[2 \pi I_0 (\beta r \sigma)\right]}. 
\end{align}

The critical temperature $T_c=1/\beta_c$ is then obtained as
\begin{align}
a_2(T=T_c)=0,    
\end{align}
yielding
\begin{align}
    \beta_c=-\frac{J}{\sigma^2}\left([r]|_{\beta_c}-\frac{[r^3]|_{\beta_c}}{2}\right).
    \label{eq:beta_c}
\end{align}

Let us now discuss the weak-disorder limit, i.e., $\sigma \to 0$. In this limit, we have on using Eq.~\eqref{eq:Bessel function expansion around zero} that
\begin{align}
    \log{[2\pi I_0(\beta r \sigma)]} \approx \log{2\pi} + \frac{1}{4} \beta^2  \sigma^2 r^2 - \frac{1}{64} \beta^4 \sigma^4 r^4,  
\end{align}
yielding $[r]=\beta^2 \sigma^2/2 - \beta^4 \sigma^4/8 + \log(2\pi)$ and $[r^3]=2 \beta^2 \sigma^2 - 3 \beta^4 \sigma^4/4 + 2 \log(2\pi)$. On using Eq.~\eqref{eq:beta_c}, we finally get $\beta_c=2/J$, the known critical temperature for the mean-field $XY$ model.

Now, we discuss the other limit $\beta \to \infty$, for which 
one has~\cite{dlmf} 
 \begin{align}
     I_0(\beta r \sigma) \approx \frac{\exp{\left(\beta r \sigma\right)}}{\sqrt{2\pi \beta r \sigma}}.
 \end{align} 
Using the above result and taking the limit $\beta \to \infty$ in Eq.~\eqref{eq:free_energy_XY_rotational}, we obtain 
\begin{align}{\label{eq:Zero_temp_XY_free_energy_Gaussian}}
    \frac{\mathcal{F}}{N} = \frac{Jm^2}{2}- \sigma \exp{\left(-\frac{J^2 m^2}{2\sigma^2}\right)}\int_{0}^{\infty} dr~r^2 \exp{\left(-\frac{r^2}{2}\right)}I_0\left(\frac{Jmr}{\sigma}\right),
\end{align}
which can be expanded around $m=0$ in the spirit of the Landau theory by using Eq.~\eqref{eq:Bessel function expansion around zero}, as
\begin{align}{\label{eq:Landau_zero_temp_XY_gaussian}}
    &\frac{\mathcal{F}}{N}= \frac{Jm^2}{2}- \sigma \int_{0}^{\infty} dr~r^2 \exp{\left(-\frac{r^2}{2}\right)}\nonumber \\
    &\times \left[1+m^2\left(-\frac{J^2}{2\sigma^2}+\frac{J^2r^2}{4\sigma^2}\right)+m^4\left(\frac{J^4}{8\sigma^4}+\frac{J^4r^4}{64\sigma^4}-\frac{J^4 r^2}{8\sigma^4}\right)\right]+\ldots\nonumber \\
    &=m^2 \left(\frac{J}{2}-\frac{J^2}{4\sigma}\sqrt{\frac{\pi}{2}}\right)+m^4\frac{J^4}{64\sigma^3}\sqrt{\frac{\pi}{2}}+\ldots.
\end{align}
By equating the coefficient of the $m^2$-term to zero, we obtain the critical disorder strength $\sigma_c = (J/2) \sqrt{\pi/2} \approx 0.6266 J$, above which the system (being at $T=0$, it would be in its ground state) shows a continuous phase transition from a low-$\sigma$ ferromagnetic to a high-$\sigma$ paramagnetic phase, as shown in Fig.~\ref{fig:XY_Gaussian_ground_state_m_vs_sigma}. Schematic free-energy landscapes for different representative values of $\sigma$, shown in Fig.~\ref{fig:fe}, clearly indicate the existence of a continuous phase transition close to $\sigma=0.6266 J$. This $\sigma_c$ also serves as an upper bound on the disorder strength above which no ferromagnetic ordering in the system survives at any temperature.

Now, we turn to the discussion of the critical temperature at finite temperatures and disorder strengths. To proceed further, for given values of $T>0$ and $\sigma>0$, we numerically integrate Eq.~\eqref{eq:free_energy_XY_rotational} to obtain the corresponding free-energy landscape as a function of $m$. We then look for the value of $0 \le m \le 1$ that minimizes the free energy for given $T$ and $\sigma$. This minimizing $m$ is the equilibrium magnetization value, and at the transition point, the equilibrium value becomes zero. From Fig.~\ref{fig:XY_Gaussian_m_vs_T_theory}, it is evident that the system undergoes a continuous phase transition in $m$ versus $T$, just as in the case without disorder. However, the presence of disorder manifests itself by monotonically shrinking the temperature range over which the system remains ordered as the disorder strength increases.

\subsubsection{Bimodal distribution}
In this part, we discuss the case in which the disorder distribution is bimodal, Eq.~\eqref{eq:bimodal}. Equation~\eqref{eq:XY_modified_free_energy} gives 
\begin{align}{\label{eq:XY_bimodal_free_energy}}
    \frac{\mathcal{F}}{N} & = \frac{J m^2}{2} - \frac{1}{4\beta} \left[\log{(2\pi I_0^{++})} + \log{(2\pi I_0^{+-})}\right] \nonumber \\  
    &~~~~-\frac{1}{4\beta} \left[\log{(2\pi I_0^{-+})} + \log{(2\pi I_0^{--})} \right].
\end{align}
with $I_0^{+ + }$, $I_0^{+ -}$, $I_0^{-+}$, $I_0^{--}$ denoting the cases $\{h^x=h_0, h^y=h_0\}$,  $\{h^x=h_0, h^y=-h_0\}$,  $\{h^x=-h_0, h^y=h_0\}$, and $\{h^x=-h_0, h^y=-h_0\}$, respectively; for example, we have $I_0^{+-} \equiv I_0 \left(\beta\sqrt{(h_0 + Jm_x)^2 + (-h_0 + Jm_y)^2} \right)$.
Clearly, the bimodal distribution does not possess rotational symmetry, as in the case of Gaussian. This does not allow the free energy per spin in Eq.~\eqref{eq:XY_bimodal_free_energy} to be written only in terms of $m = \sqrt{m_x^2 + m_y^2}$. In passing, note that the expression in Eq.~\eqref{eq:XY_bimodal_free_energy} remains invariant under the following transformations:
\begin{align}{\label{eq:symmetries_XY_bimodal}}
\begin{aligned}
    &(i) \quad m_x \to m_y, \quad m_y \to m_x, \\
    &(ii) \quad m_x \to -m_y, \quad m_y \to m_x, \\
    &(iii) \quad m_x \to -m_y, \quad m_y \to -m_x, \\
    &(iv) \quad m_x \to m_y, \quad m_y \to -m_x, \\
    &(v) \quad m_x \to -m_y, \quad m_y \to -m_x.
\end{aligned}
\end{align}

Let us first discuss the zero-temperature limit, i.e., the limit $\beta \to \infty$. Using $\lim_{\beta \to \infty} I_0(\beta x) = \exp{\left(\beta x\right)}/{\sqrt{2\pi\beta x}}$~\cite{dlmf}, one gets
\begin{align}{\label{eq:XY_bimodal_ground_state_free_energy}}
    \frac{\mathcal{F}}{N} & = \frac{J m^2}{2} - \frac{1}{4} \left[ f^{++} + f^{+-} + f^{-+} + f^{--}  \right],  
\end{align}
with 
\begin{align}
f^{++} &\equiv  \sqrt{(h_0 + J m_x)^2 + (h_0 + J m_y)^2}, \\
f^{+-} &\equiv \sqrt{(h_0 + J m_x)^2 + (-h_0 + J m_y)^2}, \\
f^{-+} &\equiv \sqrt{(-h_0 + J m_x)^2 + (h_0 + J m_y)^2}, \\
f^{--} &\equiv \sqrt{(-h_0 + J m_x)^2 + (-h_0 + J m_y)^2}.
\end{align}
It is worth mentioning that the symmetries mentioned in Eq.~\eqref{eq:symmetries_XY_bimodal} also hold for the zero-temperature free energy (the ground-state free energy) in Eq.~\eqref{eq:XY_bimodal_ground_state_free_energy}. These symmetries manifest themselves in the landscape of the ground-state free energy as a function of $(m_x,m_y)$ for all values of $h_0$, as evident from Fig.~\ref{fig:XY_bimodal_ground_state}, where we see that the white dashed lines divide the landscape into regions in which the free energy has identical behavior. More precisely, there is discrete rotational symmetry present in the landscape, as opposed to the continuous rotational symmetry present when the disorder distribution is Gaussian.   

To obtain the ground-state magnetization $m\equiv\sqrt{m_x^2 + m_y^2}$ from the zero-temperature free-energy landscape, we use the discrete rotational symmetries mentioned above and numerically extract the minimizing $(m_x,m_y)$ coordinates by restricting ourselves only to the region bounded by the two orange lines in Fig.~\ref{fig:XY_bimodal_ground_state}, together with the constraint $\sqrt{m_x^2 + m_y^2} \leq 1$. We find that the ground-state magnetization undergoes a first-order transition between ferromagnetic and paramagnetic phases with increasing disorder strength $h_0$, captured by the sudden jump of magnetization at $h_0^c\approx 0.4422J$ in Fig.~\ref{fig:XY_bimodal_ground_m_vs_h0}. 

\begin{figure}
    \centering
    \includegraphics[width=1\linewidth]{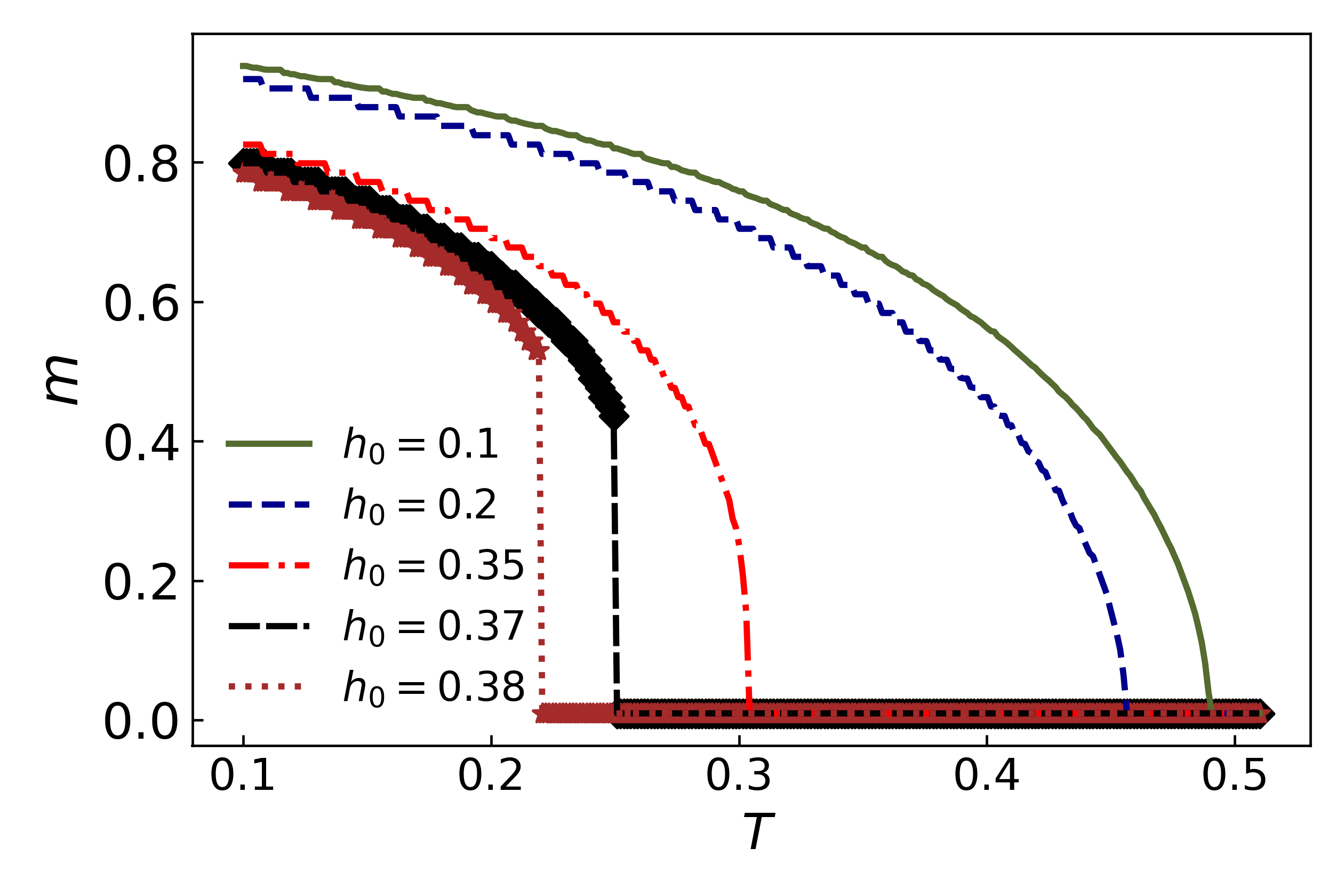}
    \caption{For the $n=2$ case in Eq.~\eqref{eq:H}, corresponding to the $XY$ spin model with bimodal disorder distribution~\eqref{eq:bimodal}, the variation of equilibrium magnetization $m$ with temperature $T$ is shown for various disorder strengths $h_0$. We consider here $J=1$. The data in this figure are generated by numerically extracting the global minimum of Eq.~\eqref{eq:XY_bimodal_free_energy}.}
    \label{fig:XY_bimodal_m_vs_T_theory}
\end{figure}

\begin{figure}
    \centering
    \includegraphics[width=1.0\linewidth]{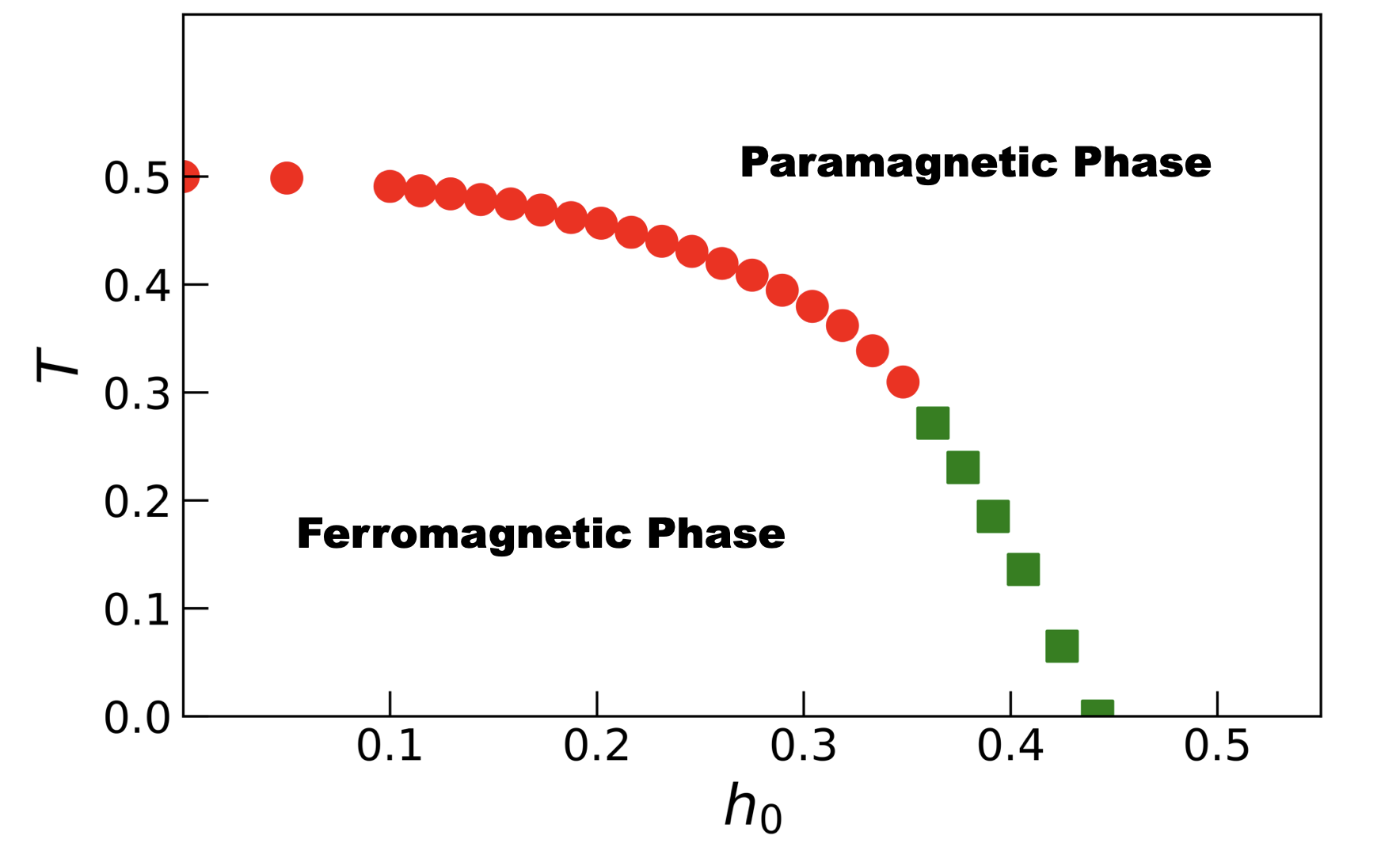}
    \caption{Phase diagram of the model~\eqref{eq:H} for $n=2$ and with the bimodal distribution~\eqref{eq:bimodal} for the disordered fields. The green and the red symbols denote respectively first-order and continuous transitions. We consider here $J=1$. The data are based on our analytical results, Eq.~\eqref{eq:XY_bimodal_free_energy}. Numerical estimation of the tricritical point, defined as the point where the first-order and continuous transition lines meet each other, is discussed in Appendix~\ref{apperror}.}
    \label{fig:n2-bimodal-phase-diag}
\end{figure}

Let us now discuss the finite-temperature equilibrium properties of the system. In Fig.~\ref{fig:XY_bimodal_finite_temp_free_energy_landscape}, we show the free-energy behavior in the $(m_x,m_y)$-plane for representative values of $h_0$ and $T$. To obtain the equilibrium magnetization at finite temperatures, we follow the same procedure as followed above. Namely, using the symmetries mentioned in Eq.~\eqref{eq:symmetries_XY_bimodal}, we find the equilibrium magnetization $m=\sqrt{m_x^2 + m_y^2}$ by extracting the  $(m_x,m_y)$-coordinates that minimize the free energy in Eq.~\eqref{eq:XY_bimodal_free_energy}. In Fig~\ref{fig:XY_bimodal_m_vs_T_theory}, we show the behavior of equilibrium magnetization $m$ versus $T$, for several values of $h_0$. One observes that with increasing $h_0$, a transition that was continuous becomes first order, indicated respectively by the absence and presence of a jump in the value of $m$. The phase diagram in the $(h_0,T)$-plane is shown in Fig.~\ref{fig:n2-bimodal-phase-diag}, in which one may observe a continuous and a first-order transition line between a ferromagnetic and a paramagnetic phase joining at a tricritical point $(h_0^\mathrm{tri},T^\mathrm{tri})\approx (0.3512J,0.3006J) $.   
\section{Phase transitions in the $n=3$ case}
\label{sec:n3}
We now consider the effect of disorder on the critical point for the case $n=3$, i.e., the case of classical Heisenberg spins. The spin components for this case can be written as $(S^x,S^y,S^z) = (\sin{\theta}\cos{\phi},\sin{\theta}\sin{\phi},\cos{\theta})$, with $0\leq \theta \leq \pi$ and $0 \leq \phi < 2\pi$. The free energy per spin for this case is obtained from Eq.~\eqref{eq_F} as 
\begin{align}{\label{eq:Heisenberg_free_energy}}
\frac{\mathcal{F}}{N} &= \frac{J(m_x^2 + m_y^2 + m_z^2 )}{2} \nonumber \\
&~~~~-\frac{1}{\beta} \int P(\Vec{h}) d\Vec{h} \log\left[\int d\Vec{S}~ \exp\left(\beta \Vec{S} \cdot (\Vec{h} + J\Vec{m})\right)\right],
\end{align}
with $\Vec{m} = (m_x,m_y,m_z)$.\\
Using the identity 
\begin{align}{\label{eq:hyperbolic_identity}} 
   & \int_{0}^\pi d\theta \sin{\theta}\int_{0}^{2\pi} \frac{d\phi}{2 \pi} \exp{(a \sin{\theta} \cos{\phi} + b \sin{\theta} \sin{\phi} + c \cos{\theta})} \nonumber \\
   & =  \frac{2 \sinh{\left( \sqrt{a^2 + b^2 + c^2} \right)}}{\sqrt{a^2 + b^2 + c^2}},
\end{align}
we get
 \begin{align}{\label{eq:free_energy_Heisenberg}}
\frac{\mathcal{F}}{N} &= \frac{J(m_x^2 + m_y^2 + m_z^2 )}{2}\nonumber \\
&~~~~-\frac{1}{\beta} \int P(\Vec{h}) d\Vec{h} \log\left[ 4 \pi \frac{\sinh{\left( \beta \sqrt{\sum_{\alpha=x,y,x} (h^{(\alpha)} + Jm_{\alpha})^2} \right)}}{\beta \sqrt{\sum_{\alpha=x,y,x} (h^{(\alpha)} + J m_{\alpha})^2 }} \right].    
\end{align}
As for $n=2$, we consider now the distributions~\eqref{eq:Gaussian} and~\eqref{eq:bimodal} for the disordered fields.\\

\subsubsection{Gaussian distribution}
Using Eq.~\eqref{eq:Gaussian}, we get 
\begin{align}
    P(\Vec{h}) = \frac{1}{(2 \pi \sigma^2)^{\frac{3}{2}}}~ \exp{\left( -\frac{{h^x}^2 + {h^y}^2 + {h^z}^2}{2 \sigma^2}
  \right)},
\end{align}
with $-\infty \leq h^x,h^y, h^z \leq \infty$. Proceeding as for the case $n=2$, we define $v_x \equiv h^x + Jm_x,~ v_y \equiv h^y + Jm_y,~v^z \equiv h^z + Jm^z $, and subsequently move to the spherical polar coordinates: $(v_x,v_y,v_z)= (r \sin{\Tilde{\theta}} \cos{\Tilde{\phi}}, r\sin{\Tilde{\theta}}\sin{\Tilde{\phi}},r\cos{\Tilde{\theta}})$, with $r \in [0,\infty)$, $0\leq \Tilde{\theta} \leq \pi$, $0\leq\Tilde{\phi}<2 \pi $. Then, under the rescaling $r \to r/\sigma$, we obtain from Eq.~\eqref{eq:free_energy_Heisenberg} the rotationally-invariant free energy per spin as 
\begin{align}{\label{eq:Heisenberg_free_energy_rotational}}
    \frac{\mathcal{F}}{N} & = \frac{J m^2}{2} - \frac{\sqrt{\frac{2}{\pi}}}{\beta}\exp{\left(-\frac{J^2 m^2}{2\sigma^2}\right)}\int_{0}^{\infty} r^2~dr~ \exp{\left(-\frac{r^2}{2}\right)}\nonumber \\
    &~~~~\times \log{\left[ \frac{4\pi \sinh{(\beta r \sigma)}}{\beta r \sigma} \right]}~\frac{\sinh{(\frac{J rm}{\sigma})}}{\frac{J r m}{\sigma}},
\end{align}
with $m= \sqrt{m_x^2 + m_y^2 + m_z^2}$.

Following the standard prescription of Landau theory, we expand the free-energy expression in Eq.~\eqref{eq:Heisenberg_free_energy_rotational} around $m=0$. Using for small $m$ that
\begin{align}
    \frac{\sinh{\left(\frac{Jrm}{\sigma}\right)}}{\frac{Jrm}{\sigma}} = 1+\frac{J^2 m^2 r^2}{6 \sigma ^2}+\frac{J^4 m^4 r^4}{120 \sigma ^4} + \ldots,
\end{align}
we obtain 
\begin{align}
    \frac{\mathcal{F}}{N} & = \frac{J m^2}{2} - \frac{\sqrt{\frac{2}{\pi}}}{\beta}\int_{0}^{\infty} r^2~dr~ \exp{\left(-\frac{r^2}{2}\right)}\log{\left[ \frac{4\pi \sinh{(\beta r \sigma)}}{\beta r \sigma} \right]} \nonumber \\
    &\times \left[ 1 - \frac{J^2m^2}{2\sigma^2} + \frac{J^4m^4}{8\sigma^4} + \ldots \right]\left[ 1+\frac{J^2 m^2 r^2}{6 \sigma ^2}+\frac{J^4 m^4 r^4}{120 \sigma ^4} + \ldots  \right]\\
    & = -\frac{\sqrt{\frac{2}{\pi}} [r^2]}{\beta} + m^2 \left( \frac{J}{2} -  \frac{\sqrt{\frac{2}{\pi}}}{\beta} \frac{J^2 [r^4]}{6 \sigma ^2}+\frac{\sqrt{\frac{2}{\pi}}}{\beta}\frac{J^2 [r^2]}{2 \sigma ^2}  + \ldots \right) \nonumber \\
    & + m^4 \left(-\frac{J^4 [r^6]}{60 \sqrt{2} \sqrt{\pi } \beta  \sigma ^4}+\frac{J^4 [r^4]}{6 \sqrt{2 \pi } \beta  \sigma ^4}-\frac{J^4 [r^2]}{4 \sqrt{2} \sqrt{\pi } \beta  \sigma ^4} + \ldots\right) \\
    & = a_o(T) + m^2 a_2(T) + m^4 a_4(T) + \ldots,
\end{align}
where we have used the definition 
\begin{align}
    [r^k] \equiv \int_{0}^\infty \mathrm{d}r~\exp{\left(-\frac{r^2}{2}\right)}\log{\left[ \frac{4\pi \sinh{(\beta r \sigma)}}{\beta r \sigma} \right]}.
\end{align}
The critical temperature $T_c=1/\beta_c$ is then obtained as 
\begin{align}
    a_2(T=T_c) = 0,
\end{align}
yielding 
\begin{align}\label{eq:beta_c_heisenberg}
    \beta_c = \frac{\sqrt{\frac{2}{\pi}} J}{\sigma^2 }\Big( [r^2] - \frac{[r^4]}{3} \Big). 
\end{align}

Let us now discuss the limit $\sigma \to 0$. In this limit, we have 
\begin{align}
    \log{\left[ \frac{4\pi \sinh{(\beta r \sigma)}}{\beta r \sigma} \right]} \approx \log(4\pi) +\frac{1}{6} \beta ^2 r^2 \sigma ^2-\frac{1}{180} \sigma ^4 \left(\beta ^4 r^4\right), 
\end{align}
yielding 
$[r^2]=(1/12
)\sqrt{\pi/2} \left(-\beta ^4 \sigma ^4+6 \beta ^2 \sigma ^2+12 \log (4 \pi )\right) $, and $[r^4] =(1/12
)\sqrt{\pi/2}\left(-7 \beta ^4 \sigma ^4 +30 \beta ^2 \sigma ^2 +36 \log (4 \pi )\right)$. On using Eq.~\eqref{eq:beta_c_heisenberg}, we get $\beta_c = 3/J$, the known critical temperature for the mean-field Heisenberg model.

\begin{figure}
    \centering
    \includegraphics[width=1\linewidth]{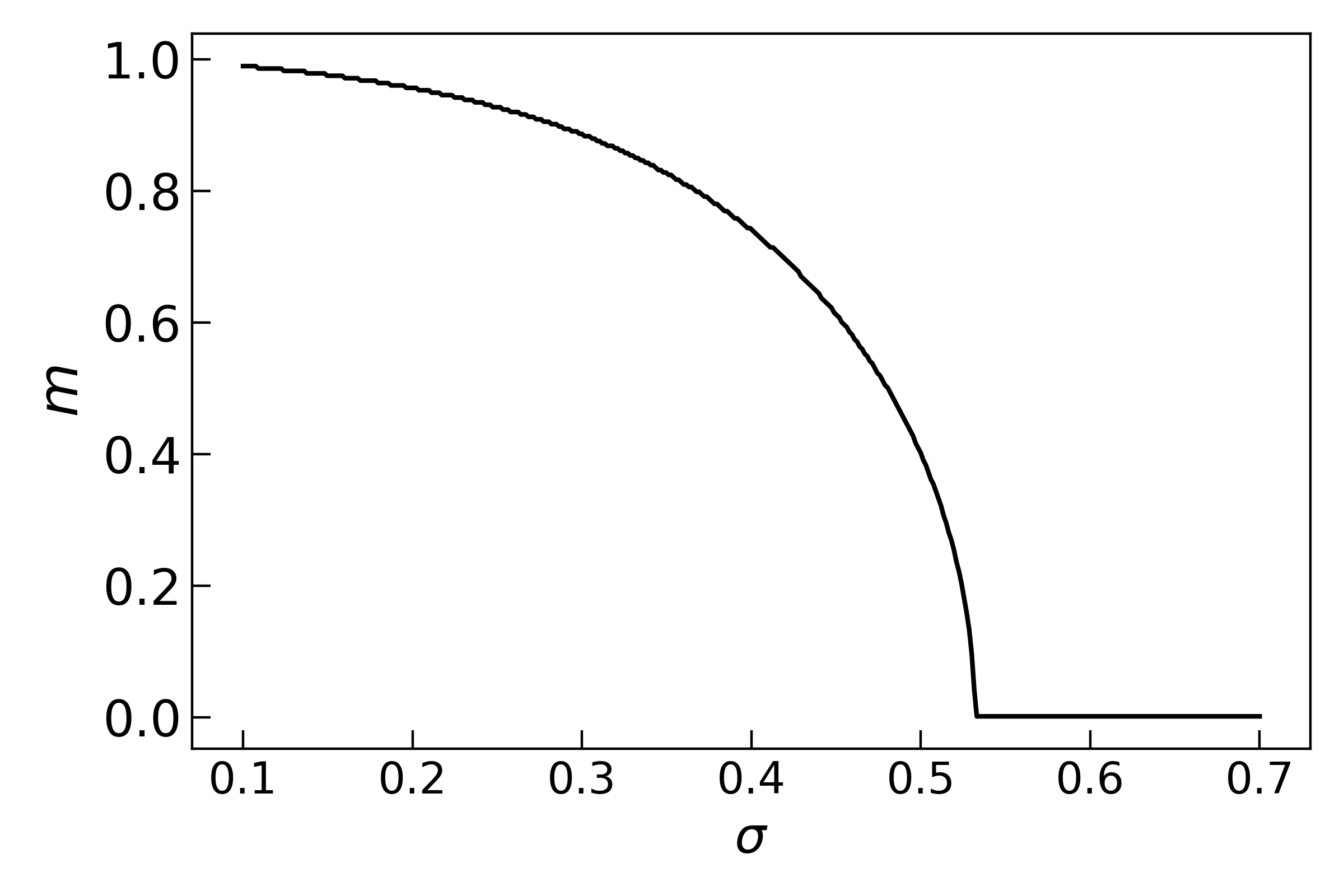}
    \caption{For the case $n=3$ in Eq.~\eqref{eq:H}, corresponding to the Heisenberg spin model with a Gaussian disorder distribution, the system at $T=0$ undergoes a continuous phase transition in equilibrium magnetization $m$ versus disorder strength $\sigma$. The critical disorder strength, identified as the value of $\sigma$ at and above which the magnetization value is zero,  is given by $\sigma_c \approx 0.5319$ (we consider here $J=1$). The data in this figure are generated by numerically extracting the global minimum of Eq.~\eqref{eq:Heisenberg-Gaussian-ground state-free energy}.}
    \label{fig:Heisenberg_Gaussian_ground_state_m_vs_sig_theory}
\end{figure}

Now, we discuss the other limit $\beta \to \infty$, for which one has $\sinh{(\beta r \sigma)} \approx \exp{\left(\beta r \sigma\right)}$. Using this asymptotic behavior and taking the limit $\beta \to \infty$ in Eq.~\eqref{eq:Heisenberg_free_energy_rotational}, we obtain
\begin{align} \label{eq:Heisenberg-Gaussian-ground state-free energy}
      &\frac{\mathcal{F}}{N} =\frac{J m^2}{2} - \sigma^2\frac{ \sqrt{\frac{2}{\pi}}~\exp{\left(-\frac{J^2 m^2}{2\sigma^2}\right)}}{J m}\nonumber \\
      &\quad \times \int_{0}^{\infty} r^2~dr~ \exp{\left(-\frac{r^2}{2}\right)} \sinh{\left(\frac{Jrm}{\sigma}\right)}, 
\end{align}
which can be expanded around $m=0$ in the spirit of the Landau theory as
\begin{align}{\label{eq:Landau_zero_temp_Heisenberg_gaussian}}
    \frac{F}{N} &= -2 \sqrt{\frac{2}{\pi }} \sigma +m^2 \left(\frac{J}{2}-\frac{\sqrt{\frac{2}{\pi }} J^2}{3 \sigma }\right) \nonumber \\ 
    & +m^4 \left(\frac{4 \sqrt{\frac{2}{\pi }} J^4}{15 \sigma ^3}-\frac{J^4}{2 \sqrt{2 \pi } \sigma ^3}\right) + m^6 \left(\frac{\sqrt{\frac{2}{\pi }} J^6}{5 \sigma ^5}-\frac{J^6}{4 \sqrt{2 \pi } \sigma ^5}\right).
\end{align}
By equating the coefficient of the $m^2$-term to zero, we obtain the critical disorder strength $\sigma_c = (2/3) \sqrt{2/\pi} J = 0.5319 J$, above which the system (being at $T=0$, it would be in its ground state) shows a continuous phase transition from a low-$\sigma$ ferromagnetic to a high-$\sigma$ paramagnetic phase, as shown in Fig.~\ref{fig:Heisenberg_Gaussian_ground_state_m_vs_sig_theory}. This $\sigma_c$ may also be interpreted as an upper bound on the disorder strength above which no ferromagnetic ordering in the system survives at any temperature.

\begin{figure}
    \centering
    \includegraphics[width=1.0\linewidth]{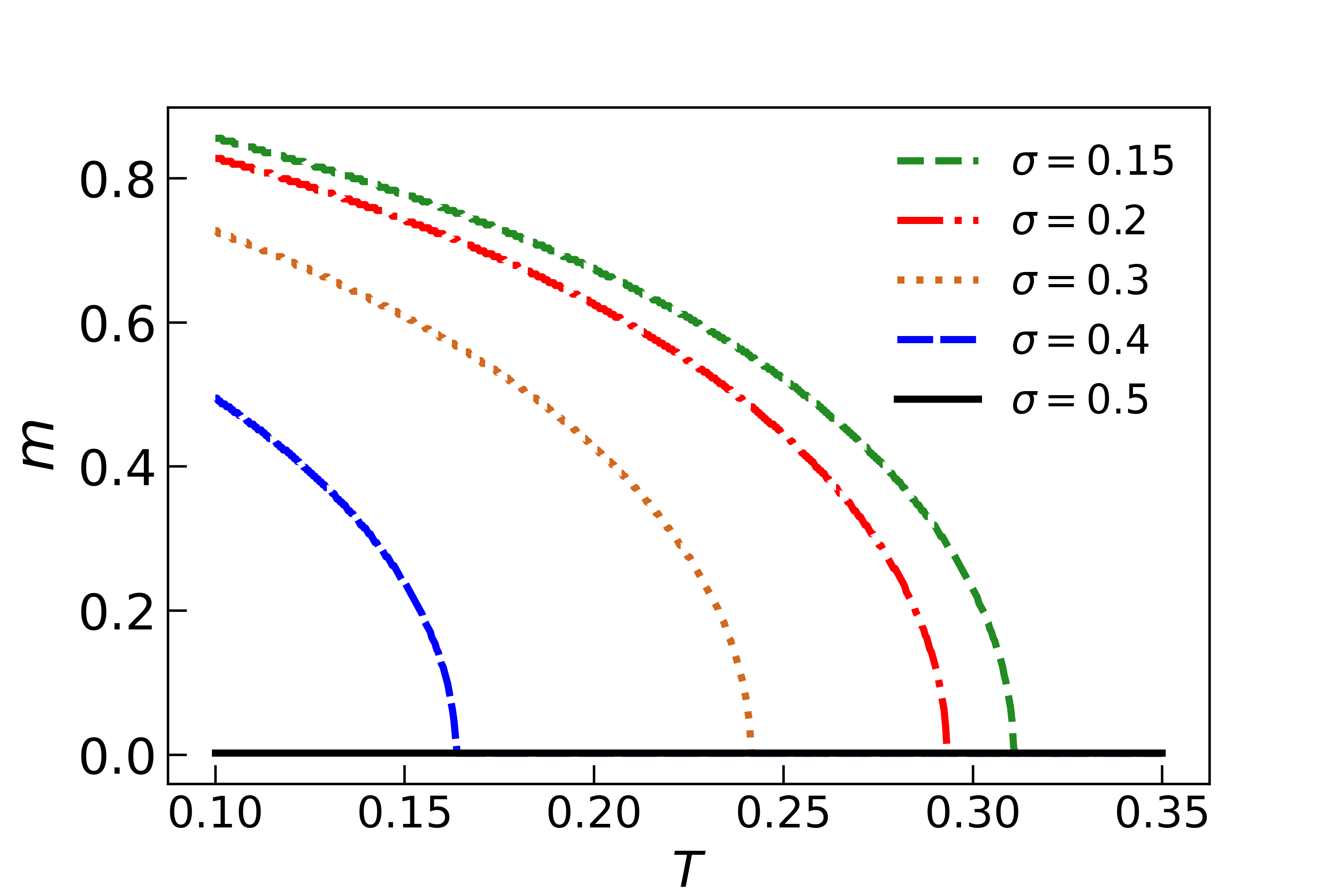}
    \caption{For the $n=3$ case in Eq.~\eqref{eq:H}
    , i.e., for Heisenberg spins with a Gaussian disorder distribution, the variation of equilibrium magnetization as a function of temperature is shown at various disorder strength $\sigma$. We consider here $J=1$. As in the case of $XY$ spins reported previously (Fig.~\ref{fig:XY_Gaussian_m_vs_T_theory}), the figure clearly demonstrates that even with non-zero disorder, the system undergoes a continuous phase transition, similar to the case without disorder. However, as expected, the critical temperature—identified as the temperature at and above which the magnetization vanishes—decreases as the disorder strength increases. The data in this figure are generated by numerically extracting the global minimum of Eq.~\eqref{eq:Heisenberg_free_energy_rotational}.} 
    \label{fig:Heisenberg_m_vs_T_gaussian_finite_temp}
\end{figure}

Turning to the discussion of the critical behavior at finite temperatures and disorder strengths, we follow the same approach as in the case $n=2$, namely, for given values of $T>0$ and $\sigma>0$, we numerically integrate Eq.~\eqref{eq:Heisenberg_free_energy_rotational} to obtain the corresponding free-energy landscape as a function of $m$. We then look for the value of $0 \le m \le 1$ that minimizes the free energy for given $T$ and $\sigma$. This minimizing $m$ is the equilibrium magnetization value, and at the transition point, the equilibrium value becomes zero. Figure~\ref{fig:Heisenberg_m_vs_T_gaussian_finite_temp} clearly demonstrates that the system undergoes a continuous transition in $m$ versus $T$ for all values of $\sigma\le \sigma_c$, akin to $n=2$ case shown in Fig.~\ref{fig:XY_Gaussian_m_vs_T_theory}.

\subsubsection{Bimodal distribution}
For this case, the disorder distribution is given by 
\begin{align}
    P(h_x,h_y,h_z) = \prod_{\alpha=x,y,z} \frac{1}{2} \left( \delta(h^{(\alpha)} + h_0) + \delta(h^{(\alpha)} - h_0) \right).
\end{align}
Following Eq.~\eqref{eq:free_energy_Heisenberg}, the free energy per spin for this distribution becomes
\begin{align}\label{eq:Heisenberg_bimodal_free_energ}
\frac{\mathcal{F}}{N} &= \frac{J(m_x^2 + m_y^2 + m_z^2 )}{2} - \nonumber \\
&\quad \frac{1}{8\beta} \left[\log (4 \pi f^{+++}) + \log (4 \pi f^{++-}) +\ldots +  \log (4 \pi f^{---})  \right],
\end{align}
where the second term on the right hand side contains contributions from all possible combinations of $(\pm h_0,\pm h_0,\pm h_0)$, e.g.,
\begin{align}
 f^{+++} &=\frac{\sinh{\left( \beta \sqrt{ (h_0 + Jm_x)^2 + (h_0 + Jm_y)^2 + (h_0 + Jm_z)^2 } \right)}}{\beta \sqrt{(h_0 + Jm_x)^2 + (h_0 + Jm_y)^2 + (h_0 + Jm_z)^2 }},\\
 f^{+-+}&=\frac{\sinh{\left( \beta \sqrt{ (h_0 + Jm_x)^2 + (-h_0 + Jm_y)^2 + (h_0 + Jm_z)^2 } \right)}}{\beta \sqrt{(h_0 + Jm_x)^2 + (-h_0 + Jm_y)^2 + (h_0 + Jm_z)^2 }},\\
 &\vdots \nonumber \\
 f^{---}&=\frac{\sinh{\left( \beta \sqrt{ (-h_0 + Jm_x)^2 + (-h_0 + Jm_y)^2 + (-h_0 + Jm_z)^2 } \right)}}{\beta \sqrt{(-h_0 + Jm_x)^2 + (-h_0 + Jm_y)^2 + (-h_0 + Jm_z)^2 }}.
\end{align}
Clearly, as in the case of $n=2$, the free energy has discrete rotational symmetries in the $(m_x,m_y,m_z)$-plane, namely, invariance under the following transformations: 
\begin{align}{\label{eq:symmetries_Heisenberg_bimodal}}
\begin{aligned}
    &(i) ~~ m_x \to -m_x, ~~ m_y \to m_y,~~ m_z \to m_z,  \\
    &(ii) ~~ m_x \to m_x, ~~ m_y \to -m_y, ~~ m_z \to m_z,  \\
    &(iii) ~~ m_x \to m_x, ~~ m_y \to m_y, ~~ m_z \to -m_z, \\
    &(iv) ~~ m_x \to m_y, ~~ m_y \to m_y, ~~ m_z \to m_z,  \\
    &(v) ~~ m_x \to -m_y, ~~ m_y \to m_y, ~~ m_z \to m_z. 
\end{aligned}
\end{align}

\begin{figure}
    \centering
    \includegraphics[width=1\linewidth]{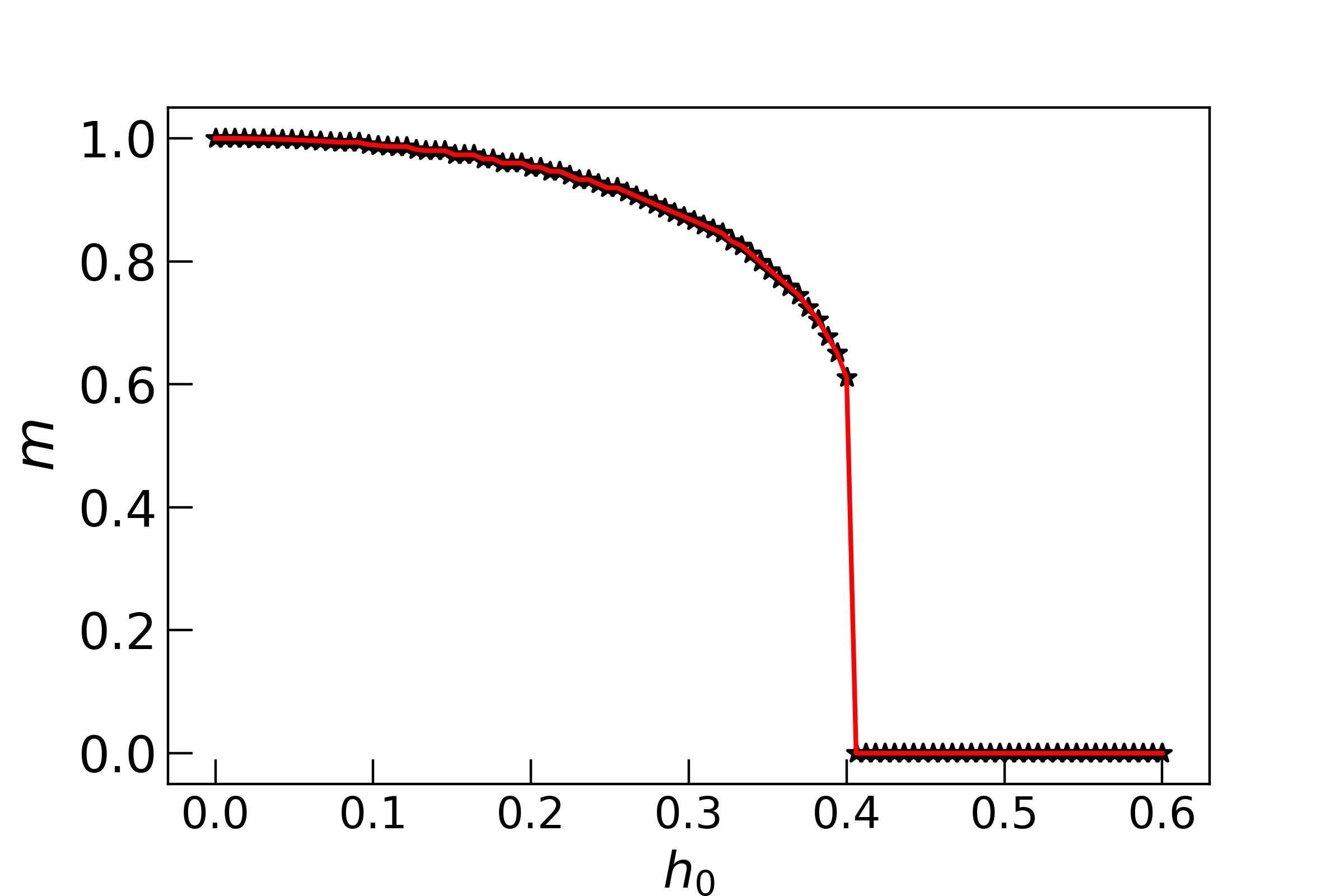}
    \caption{For the case $n=3$ in Eq.~\eqref{eq:H}, corresponding to the Heisenberg-spin model with bimodal disorder distribution, the $T=0$ equilibrium magnetization (i.e., the ground-state magnetization) of the system undergoes a first-order phase transition from a ferromagnetic to a paramagnetic phase with increasing disorder strength $h_0$, indicated by the sudden jump in magnetization $m$ at the critical value of disorder given by $h_0^c\approx 0.4060$ (We consider here $J=1$). The data in this figure are generated by numerically extracting the global minimum of Eq.~\eqref{eq:Heisenberg-ground-state-free-eenrgy}.}
    \label{fig:Heisenberg_m_vs_h0_ground_state}
\end{figure}

Let us first focus on the $\beta \to \infty$ limit, which yields $\sinh{(\beta x)}/(\beta x) \approx \exp{(\beta x)}/(2\beta x)$. Using this and taking the limit in Eq.~\eqref{eq:Heisenberg_bimodal_free_energ}, we obtain the zero-temperature free-energy per spin as 
\begin{align}\label{eq:Heisenberg-ground-state-free-eenrgy}
    \frac{\mathcal{F}}{N} &= \frac{J(m_x^2 + m_y^2 + m_z^2 )}{2} - \frac{1}{8}\left[ g^{+++} + g^{++-} + \ldots + g^{---}   \right],
\end{align}
where we have defined the quantity $g^{++-}$ as $g^{++-}\equiv \sqrt{(h_0 + Jm_x)^2 + (h_0 + Jm_y)^2 + (-h_0 + Jm_z)^2}$, and so on for the other $g$'s. Clearly, the zero-temperature free energy in Eq.~\eqref{eq:Heisenberg-ground-state-free-eenrgy} retains the symmetries in Eq.~\eqref{eq:symmetries_Heisenberg_bimodal}. Proceeding as for the case of $n=2$, we use the above symmetries and look for $(m_x,m_y,m_z)$ coordinates that minimize the zero-temperature free energy. As in the case for $n=2$, we find that the ground-state magnetization $m=\sqrt{m_x^2 + m_y^2 + m_z^2}$ undergoes a first-order transition, demonstrated by the jump of magnetization at $h_0^c\approx 0.4060J$  in Fig.~\ref{fig:Heisenberg_m_vs_h0_ground_state}.

\begin{figure}
    \centering
    \includegraphics[width=1\linewidth]{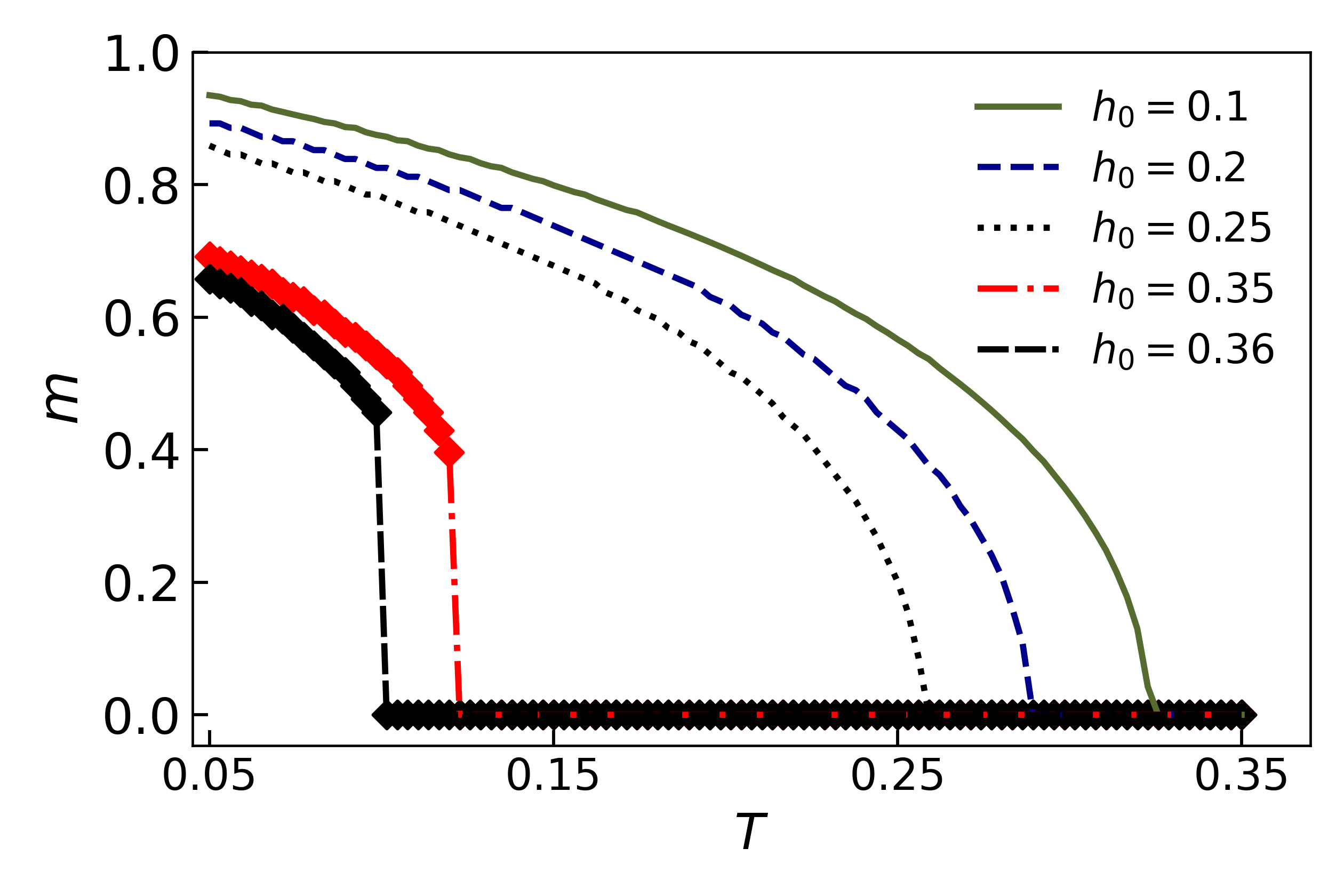}
    \caption{For the $n=3$ case in Eq.~\eqref{eq:H}, corresponding to the Heisenberg-spin model with bimodal disorder distribution~\eqref{eq:bimodal}, the variation of equilibrium magnetization $m$ with temperature $T$ is shown for various disorder strengths $h_0$. We consider here $J=1$. The data in this figure are generated by numerically extracting the global minimum of Eq.~\eqref{eq:Heisenberg_bimodal_free_energ}.}
    \label{fig:Heisenberg_bimodal_m_vs_T_theory}
\end{figure}

\begin{figure}
    \centering
    \includegraphics[width=1.0\linewidth]{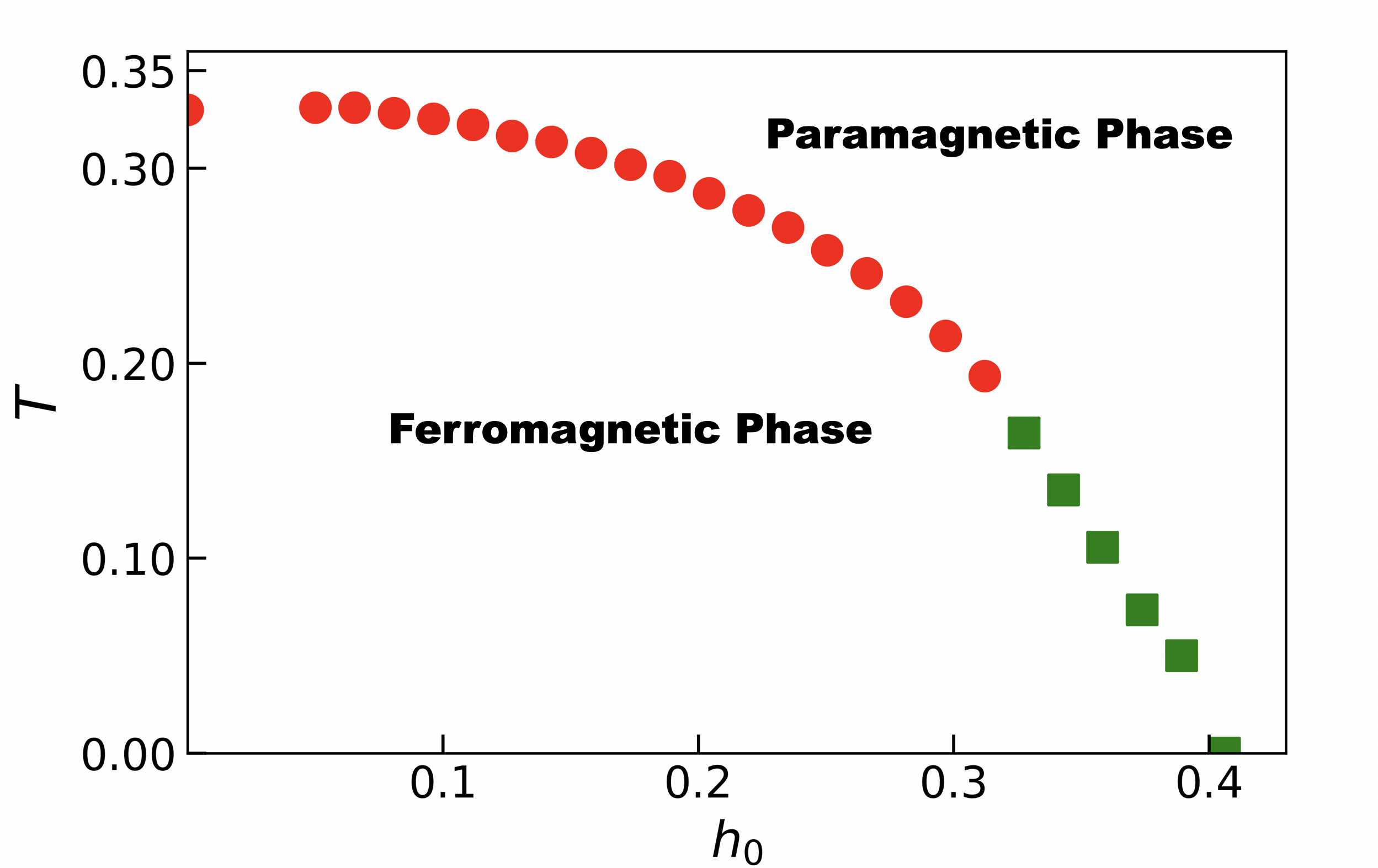}
    \caption{Phase diagram of the model~\eqref{eq:H} for $n=3$ and with the bimodal distribution~\eqref{eq:bimodal} for the disordered fields. The green and the red symbols denote respectively first-order and continuous transitions. We consider here $J=1$. The data are based on our analytical results, Eq.~\eqref{eq:Heisenberg_bimodal_free_energ}.}
    \label{fig:Heisenberg_bimodal_phase_diagram}
\end{figure}

Next, we turn to the behavior of equilibrium magnetization at finite temperatures and non-zero disorder strength $h_0$.In Fig~\ref{fig:Heisenberg_bimodal_m_vs_T_theory}, we show the behavior of equilibrium magnetization $m$ versus $T$, for several values of $h_0$. One observes that with increasing $h_0$, a transition that was continuous becomes first order, indicated respectively by the absence and presence of a jump in the value of $m$. The phase diagram in the $(h_0,T)$-plane is shown in Fig.~\ref{fig:Heisenberg_bimodal_phase_diagram}, in which one may observe a continuous and a first-order transition line between a ferromagnetic and a paramagnetic phase joining at a tricritical point $(h_0^\mathrm{tri},T^\mathrm{tri})\approx (0.3245J,0.1712J) $.

\section{Absence of self-averaging near criticality} \label{sec:absence of self-averaging}
An intriguing property prevalent in typical quenched-disordered systems is the absence of self-averaging close to a critical point, implying thereby that extensive thermodynamic variables exhibit strong sample-to-sample fluctuations near the critical point. Several detailed studies on this phenomenon for systems with disordered couplings can be found in Refs.~\cite{Aharony_Harris_self_averaging,Wiseman_self_averaging,Wiseman-PRL}. For a finite system of size $N$, the degree of self-averaging for a thermodynamic variable $X$ in equilibrium at temperature $T$ can be assessed by analyzing the quantity 
\begin{align}
R_X(T) \equiv \frac{\overline{X^2(T)} - \overline{X(T)}^2}{\overline{X(T)}^2},
\end{align}
where the overline denotes averaging over disorder realizations. Examining the exponent of power-law scaling $R_X(T) $ with $N$, $R_X(T) \sim N^{-z}$, provides insights into the self-averaging behavior. Specifically, (i) if one has $0 < z < 1$, the system is said to be weakly self-averaging; (ii) if one has $z \geq 1$, the system is strongly self-averaging; (iii) if, on the other hand, one has $z \le 0$, the system is non-self-averaging, implying that sample-to-sample fluctuations persist even in the thermodynamic limit $N\to \infty$. It is worth noting that the critical temperature $ T_c $ in a disordered system depends on specific disorder realizations, and, consequently, a detailed study of the distribution of $T_c $ is also desirable.

In the light of the above discussion, we now present a detailed numerical analysis of the self-averaging property of two thermodynamic variables, namely, the susceptibility $\chi(T)$ and the magnetization $m(T)$, for our Hamiltonian in Eq.~\eqref{eq:H} with $n = 2$ ($XY$ spins) and Gaussian random fields, based on Monte Carlo simulation results (for details on the algorithm, see Appendix~\ref{Monte-Carlo}). We consider here $J=1$. In our simulations, we use for a given realization of the disordered fields the following definition of the susceptibility:
\begin{align} \label{eq:susceptibility-linear-response}
    \chi(T) \equiv  \frac{1}{N}\beta  [\langle M^2 \rangle(0,T)- (\langle M \rangle(0,T))^2],
\end{align}
where $M \equiv \sqrt{\sum_{i=1}^N \left((S_i^x)^2 + (S_i^y)^2 \right)}$ is the un-normalized magnetization of a spin configuration; here $\langle M \rangle(h,T)$ denotes average magnetization in canonical equilibrium at temperature $T$ and in presence of a magnetic field of strength $h$, see Appendix~\ref{linear-response}. 
\begin{figure}
    \centering
    \includegraphics[width=1.0\linewidth]{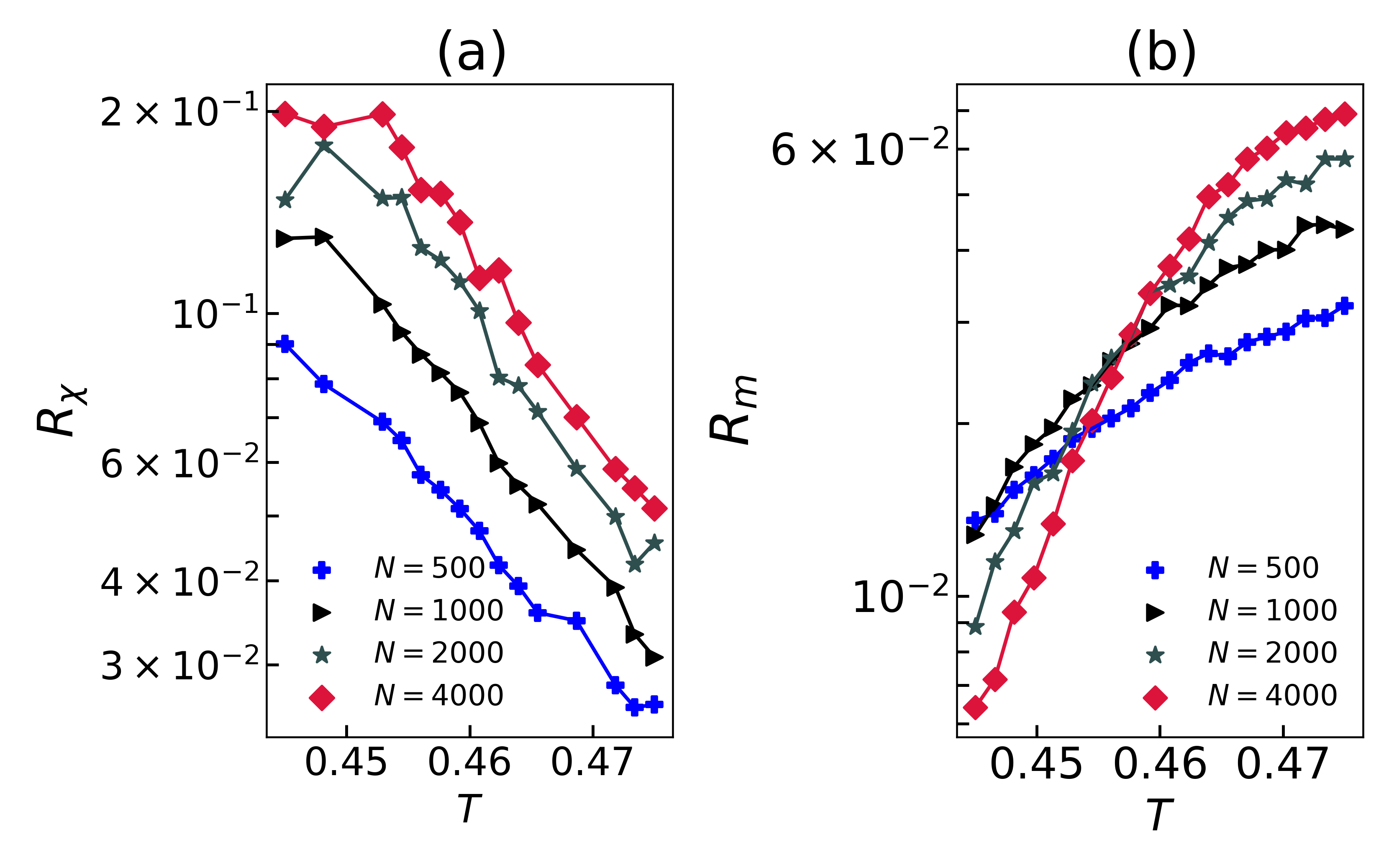}
    \caption{For the case $n=2$ in Eq.~\eqref{eq:H}, corresponding to $XY$ 
    spin model with disordered fields sampled from a Gaussian distribution with $\sigma=0.2$, we demonstrate the absence of self-averaging near the transition point $T_c=0.4607$ (we consider here $J=1$) by examining $R_{\chi}$ and $R_{m}$ as a function of $T$ for increasing system sizes. The data are generated using the Monte Carlo algorithm detailed in the text.  All data presented here are taken over 500 disorder realizations.}
    \label{fig:susceptibility-self-averaging}
\end{figure}

\begin{figure*}
    \centering
    \includegraphics[width=\linewidth]{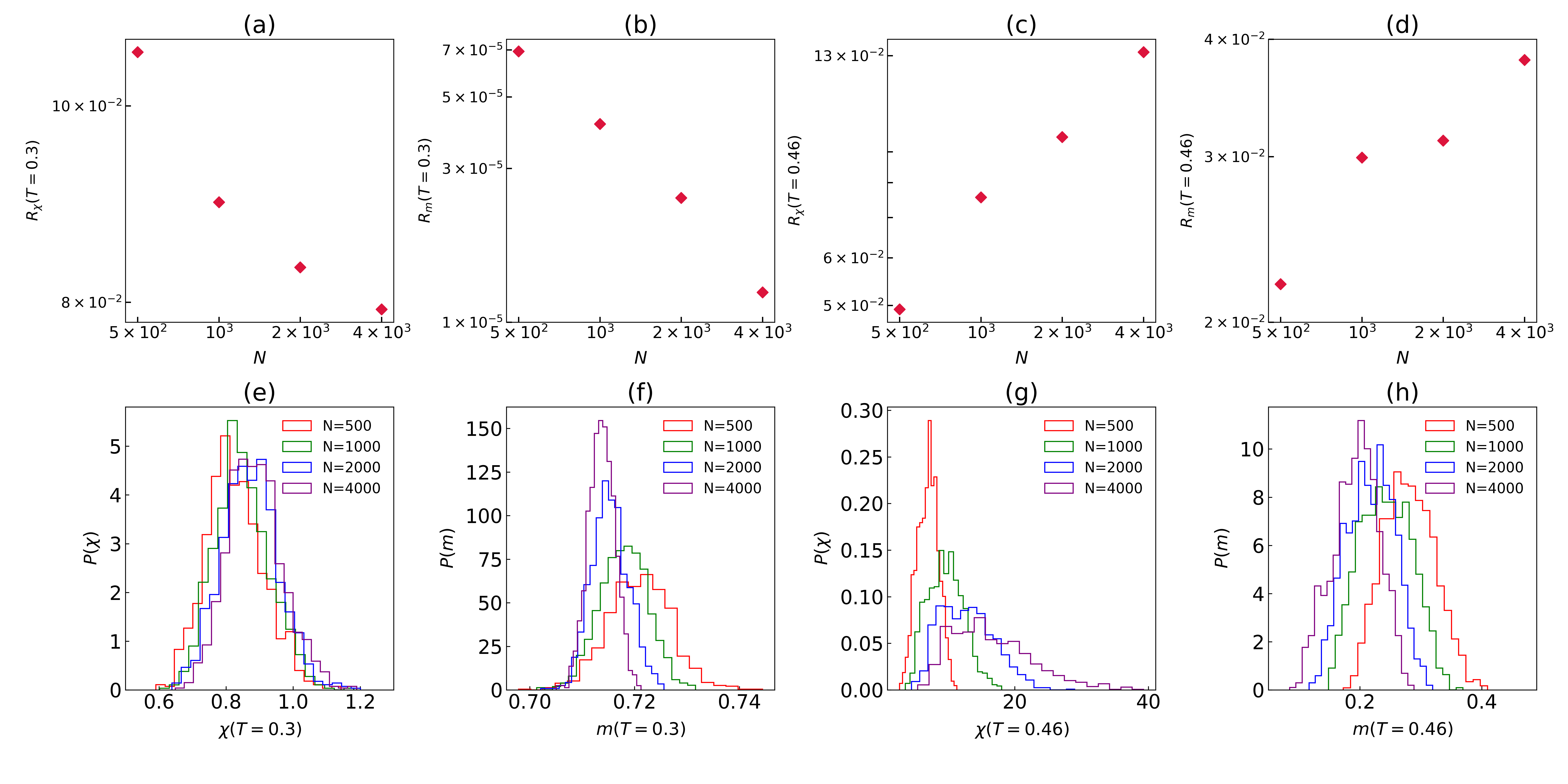}
    \caption{For the case $n=2$ in Eq.~\eqref{eq:H}, corresponding to $XY$ spin model with disordered fields sampled from a Gaussian distribution with $\sigma=0.2$, we demonstrate the presence of self-averaging for magnetization and susceptibility away from the critical point, and its absence near the critical point. Specifically, at $T=0.3$, which is away from the critical point $T_c=0.4607$, panels (a) and (b) show that $R_{\chi}$ and $R_{m}$ both decrease with increasing $N$, implying the presence of self-averaging. The corresponding histograms over disorder realizations for different $N$ are shown in panels (e) and (f), respectively. In contrast, at $T=0.46$, which is near the transition temperature $T_c$, panels (c) and (d) show that $R_{\chi}$ and $R_{m}$ increase with increasing $N$, indicating the absence of self-averaging. The corresponding histograms over disorder realizations for different $N$ are shown in panels (g) and (h), respectively. The data are generated using the Monte Carlo algorithm detailed in the text.  All data are taken over $1000$ disorder realizations. We consider here $J=1$.}
    \label{fig:XY-Self-averaging}
\end{figure*}

In Fig.~\ref{fig:susceptibility-self-averaging}, we demonstrate for disorder strength $\sigma =0.2$ how the quantities $R_{\chi}(T)$ and $R_m(T)$ behave near the theoretically-estimated transition temperature $T_c=0.4607$ (see Fig.~\ref{fig:XY_Gaussian_m_vs_T_theory}). We clearly see that near criticality, both $R_\chi(T)$ and $R_m(T)$ grow with increasing system size $N$, indicating that the system is non-self-averaging. This suggests that disorder has a crucial role to play near criticality in our system~\cite{Aharony_Harris_self_averaging}. In general, one expects that in systems showing critical behavior, self-averaging (weak or strong) property will be restored sufficiently away from the critical point. In Fig.~\ref{fig:XY-Self-averaging}, we demonstrate this fact by a comparative study of $R_{\chi}$ and $R_{m}$ and their respective histograms for two representative choices of the temperature, namely, $T=0.3$, which is away from the transition point, and the other, $T=0.46$, that is very close to transition point. For $T=0.3$, we see from Fig.~\ref{fig:XY-Self-averaging}(a),(b), that $R_{\chi}$ and $R_{m}$ both decrease with increasing $N$, implying holding of the self-averaging property. In contrast, for $T=0.46$, Fig.~\ref{fig:XY-Self-averaging}(c), (d) show that $R_{\chi}$ and $R_{m}$ increase monotonically with $N$, implying absence of the self-averaging property. This feature is also reflected in the broad nature of the susceptibility histogram as shown in Fig.~\ref{fig:XY-Self-averaging}(g).  

\begin{figure}
    \centering
    \includegraphics[width=1\linewidth]{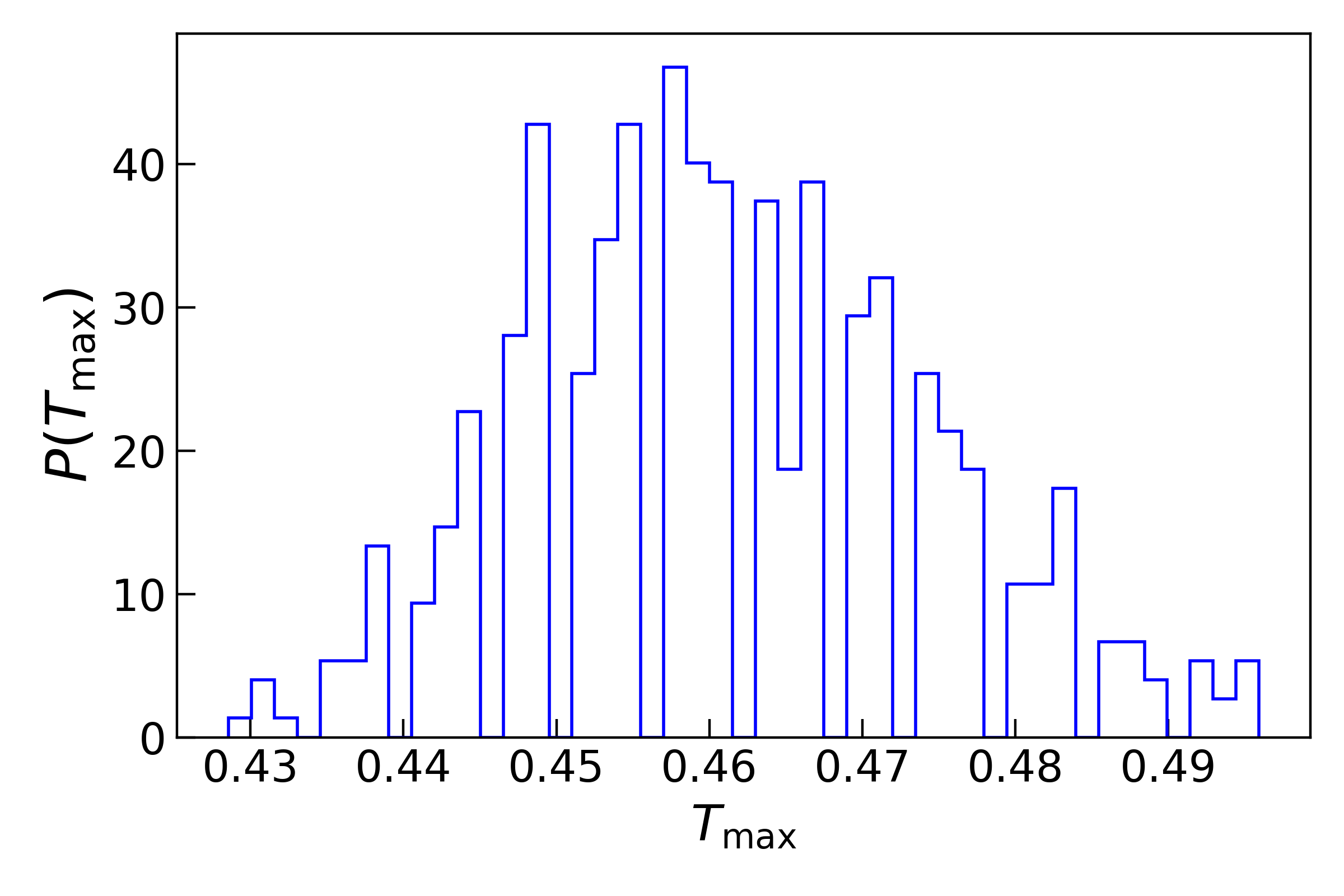}
    \caption{For the case $n=2$ in Eq.~\eqref{eq:H}, corresponding to $XY$ spin model with disordered fields sampled from a Gaussian distribution with $\sigma=0.2$, the figure shows the distribution of the quantity $T_{\mathrm{max}}$ (the temperature at which the susceptibility $\chi(T)$ attains its peak) for $N=1000$ and for $500$ disorder realizations. The data are generated using the Monte Carlo algorithm detailed in the text. We consider here $J=1$.}
    \label{fig:Tmax_histogram}
\end{figure}

Additionally, for $N =1000$ and for $500$ disorder realizations, we show in Fig.~\ref{fig:Tmax_histogram} the histogram of $T_{\mathrm{max}}$, where $T_\mathrm{max}$ is numerically extracted by identifying the temperature value at which susceptibility has a peak for a given disorder realization (indeed, a continuous phase transition is signaled by having a susceptibility that in the thermodynamic limit diverges at the critical temperature). We find that the distribution has a broad nature near the critical point, and the mean $T_{\mathrm{max}} =0.4611$ is close to the theoretical prediction for the transition point, $T_c \approx 0.4607$.

In passing, we remark that we do not anticipate significant change in the qualitative picture if one chooses a different disorder strength $\sigma$, or, a different value of $n$ than those considered above. Also, we expect similar behavior as regards holding or violation of the self-averaging property if one considers a bimodal disorder distribution with disorder strength $h_0$ restricted to the region in which the model shows a continuous phase transition.  

\section{Conclusions}
\label{sec:conclusion}
In this work, we studied the canonical-equilibrium properties of Random
Field $O(n)$ Models in presence of mean-field interactions and 
disordered fields that are either Gaussian-distributed or sampled from a symmetric bimodal distribution and acting on individual spins. On the basis of an explicit expression for the free energy per spin of the system, derived using the mean-field approximation applicable for any disorder distribution without a broad tail and the replica trick, applicable when the disordered fields are Gaussian-distributed, we studied the phase transition behavior for two representative choices of $n$, namely, $n=2$ ($XY$ spins) and $n=3$ (Heisenberg spins). For both the cases, we demonstrated that the magnetization exhibits a continuous phase transition as a function of temperature when the disorder distribution is Gaussian. Contrarily and remarkably, when the distribution is bimodal, the phase transition could be either continuous or first-order with an emergent tricriticality, which we characterize analytically. In addition to locating critical and tricritical points for different disorder distributions, our work reveals strong non-self-averaging behaviour near continuous phase transitions—an aspect that is not only nontrivial but also of independent interest in the study of disordered systems. We note that the emergence of tricriticality for a bimodal disorder distribution in the case of $O(n)$ spins can be traced back to the result in Ref.~\cite{Aharony}, where a renormalization group (RG) analysis of the Random Field Ising Model (RFIM) demonstrated that tricriticality generically arises when the disorder distribution is symmetric and possesses a minimum at zero. A similar RG treatment tailored to our model is beyond the scope of the present work and is left for future investigation.

In passing, we remark that our results are of potential experimental relevance for certain classes of layered magnets, such as $(\mathrm{C_2}\mathrm{H}_5 \mathrm{NH}_3 )_2 \mathrm{Cu}\mathrm{Cl}_4$, which is known to be effectively described by a one-dimensional continuous spin Hamiltonian with both mean-field and local interactions~\cite{Sato2004,PhysRevLett.95.264101,Dauxois_2010}. Furthermore, the development of highly-controllable cavity-QED platforms has enabled the realization of long-range interacting Hamiltonians, including the Hamiltonian Mean-Field (HMF) model involving \(XY\) spins~\cite{PhysRevLett.113.203002,PhysRevA.94.023807,PhysRevLett.117.083001}. Incorporating disorder in such setups is now within experimental reach. In view of these advances, our results can be either directly tested or taken to serve as a theoretical guide for exploring phase diagrams of experimentally-accessible long-range systems with impurities.

As follow-ups, it is of immediate interest to investigate whether the critical exponents of the bare dynamics (the one in the absence of disorder) get affected by the presence of disorder. In the context when the couplings are disordered and between nearest-neighbor spins, the celebrated Harris criterion states that when the embedding dimension $d$ of the system and the critical exponent $\nu$ of the correlation length have values such that $d\ge 2/\nu$, the critical behavior of the bare model remains unaffected by the presence of disorder. Whether such a statement also holds for the model of our study is an interesting open direction to pursue. Another issue of importance concerns the equivalence of statistical ensembles in equilibrium, which is not guaranteed when the system constituents have long-range interactions that include mean-field interactions considered in this work. This question has been thoroughly addressed for both classical and quantum long-range systems, albeit in the absence of any disorder~\cite{campa2014physics}, and hence, it would be of great interest to pursue the study of how disorder may affect the equivalence or inequivalence of equilibrium ensembles for long-range systems. 

Another interesting follow-up would be to understand the emergence of long-range order in continuous spin systems beyond the Imry-Ma argument~\cite{Imry-Ma-RFIM}, which rules out such order below dimension $ d \le 4 $ for nearest-neighbour Hamiltonians with random fields. A natural extension would be to consider the class of $ O(n) $ spin models on finite-dimensional lattices, where the interactions decay with distance as a power law, $ \sim 1/r^{\alpha},~\alpha>0 $, and investigate whether long-range order can emerge under such conditions, and if so, over what range of $ \alpha $ the results predicted here for mean-field Hamiltonians apply. We conclude by noting that a recent study along these lines was conducted in Ref.~\cite{PhysRevE.108.014116} for a one-dimensional power-law diluted Hamiltonian involving $XY$ spins and random fields. Extending such analyses to higher $ n$ and to dimensions $ d = 2, 3 $ presents a promising direction for future research.

\section{Acknowledgments} We acknowledge the use of computational resources of the Department of Theoretical Physics, TIFR, and the assistance of Kapil Ghadiali and Ajay Salve. We also acknowledge the financial support of the Department of Atomic Energy, Government of India, under Project Identification No. RTI 4002. S.G. thanks ICTP–Abdus Salam International Centre for Theoretical Physics, Trieste, Italy, for support under its Regular Associateship scheme, and for hospitality in March 2025 when the paper was finalized.

\appendix

\section{Monte Carlo algorithm for $XY$ spins and Gaussian random fields} \label{Monte-Carlo}
The Monte Carlo algorithm pertaining to the study of the absence of self-averaging near criticality in Section.~\ref{sec:absence of self-averaging} is detailed below. The algorithm, modeled after the celebrated Glauber dynamics for Ising spins~\cite{glauber1963time}, aims to obtain the equilibrium state of the system at a given inverse temperature $\beta=1/T$ and for a given disorder realization~$\{\vec{h}_i\}$. 
\begin{enumerate}[label=\roman*., itemsep=0pt, topsep=0pt]
\item One prepares an initial spin configuration in which for every spin $\vec{S}_i$, the corresponding angle $\theta_i$ is independently and uniformly sampled in $[0,2\pi)$.
\item Next, we sample the disordered fields $
\{\vec{h}_i\}$ using the Gaussian distribution~\eqref{eq:Gaussian}.
\item Updating of configurations involves selecting a spin at random and attempting to change its orientation, i.e., change its $\theta$ to a new value sampled uniformly in $[0,2\pi)$. Following this, one computes the change in the energy of the system, $\Delta E$, due to the aforementioned attempted change.
\item If the energy decreases, i.e., $\Delta E<0$, the change is accepted with probability $1$ and otherwise with a reduced probability $p=\exp(-\beta \Delta E)$. 
\item $N$ attempted changes of the spin orientation constitute a single Monte Carlo time step.
\item On attaining equilibrium, physical quantities like magnetization, specific heat, etc., are measured.  
\end{enumerate}

\section{Derivation of Eq.~\eqref{eq:susceptibility-linear-response}} \label{linear-response}
The zero-field susceptibility of a magnetic system in equilibrium at inverse temperature $\beta=1/T$ is defined as  
\begin{align}
    \chi(T) \equiv \frac{\partial m(h,T)}{\partial h} \biggr|_{h\to 0}, 
\end{align}  
which quantifies the response of the system to an external magnetic field of vanishing strength $h$; here, $m(h,T)$ is the average magnetization (per degree of freedom) of the system in equilibrium at inverse temperature $\beta$ and in presence of a magnetic field of strength $h$.  
 
For concreteness, consider a generic magnetic system comprising $N$ discrete spins that we denote by $s_i;~i=1,2,\ldots,N$. In presence of an external magnetic field of strength $h$, the system is described by the Hamiltonian $H=H(C)$, with 
\begin{align}
    H = H_0 - h M, 
\end{align}  
where $C \equiv \{s_i\}$ denotes a microscopic configuration of the system, $H_0=H_0(C)$ is the Hamiltonian of the system in the absence of the field, while $M \equiv \sum_{i=1}^N s_i=M(C)$ represents the (un-normalized) magnetization of the system. When in canonical equilibrium at inverse temperature $\beta$, the average magnetization (per spin) of the system is $m = \langle M \rangle/N=m(h,T)$, where the angular brackets denote averaging with respect to the canonical equilibrium measure $P_\mathrm{eq}(C)\propto \exp{\left(-\beta H(C)\right)}$. We therefore have that 
 \begin{align}\label{eq:Mh}
    \langle M \rangle=\langle M\rangle(h,T) =\frac{\mathrm{Tr}\left[ M \exp{\left(-\beta H_0 + \beta h M\right)}  \right]}{Z(h,T)},  
\end{align}
with $\mathrm{Tr}=\sum_{\{s_i\}}$, while $Z(h,T) = \mathrm{Tr}\left[ \exp{\left(-\beta H_0 + \beta h M\right)} \right]$ denotes the canonical partition function. For small $h$, one may expand $Z(h,T)$ up to linear order in $h$, as 
\begin{align}\label{eq:Zh-linearized}
    Z(h,T) = Z(0,T)(1 + \beta h \langle M \rangle(0,T)),
\end{align}
where $Z(0,T) = \mathrm{Tr}\left[\exp{\left(-\beta H_0\right)} \right]$ refers to the canonical partition function of the system in the absence of the external field. 
Now, in this small-$h$ limit, expanding the exponential in Eq.~\eqref{eq:Mh}, we obtain to linear order in $h$ that
\begin{align}
    \langle M \rangle(h,T) & \approx \frac{\mathrm{Tr}\left[ M \exp{\left(-\beta H_0\right)}\right] + \beta h ~\mathrm{Tr}\left[  M^2 \exp{\left(-\beta H_0\right)}  \right]}{Z(0,T)(1 + \beta h \langle M \rangle(0,T))}\nonumber \\
    & \approx \langle M \rangle(0,T)+ \beta h (\langle M^2 \rangle(0,T) - (\langle M \rangle(0,T))^2).
\end{align}
Consequently, within the considered linear response, one gets 
\begin{align}\label{eq:derivMh}
    \frac{\partial \langle M \rangle(h,T)}{\partial h } = \beta [\langle M^2 \rangle(0,T) - (\langle M \rangle(0,T))^2],
\end{align}
which directly leads us to the zero-field susceptibility to be given by  
\begin{align}
    \chi(T)=\frac{1}{N} \beta [\langle M^2 \rangle(0,T) - (\langle M \rangle(0,T))^2].
\end{align}
Hence, to compute the zero-field susceptibility at a given temperature $T$, one has to just compute the standard deviation of the magnetization in canonical equilibrium at temperature $T$ and in the absence of the magnetic field~\cite{kubo1966fluctuation}. The above consideration is very general and applies to any magnetic system, including the ones studied in the main text.

\section{Error estimation in determining the transition point in Figs.~\ref{fig:XY_Gaussian_ground_state_m_vs_sigma} and~\ref{fig:XY_bimodal_ground_m_vs_h0}, and the tricritical point in Fig.~\ref{fig:n2-bimodal-phase-diag}}
\label{apperror}
To estimate the critical point $\sigma_c$ shown in Fig.~\ref{fig:XY_Gaussian_ground_state_m_vs_sigma}, we (i) obtain for every value of $\sigma$ the equilibrium magnetization by minimizing the zero-temperature free energy density given in Eq.~(\ref{eq:Zero_temp_XY_free_energy_Gaussian}) with respect to $m$ in the range $0 \leq m \leq 1$, and (ii) estimate $\sigma_c$ as the value of $\sigma$ beyond which the equilibrium magnetization has a value $\le $ a tolerance value $\epsilon$ (ideally, the equilibrium magnetization should be identicaly zero for $\sigma \ge \sigma_c$). The resulting estimates are reported in Table~\ref{tab:1}.
\begin{table}[h]
    \centering
    \renewcommand{\arraystretch}{1.8} 
    \begin{tabular}{|c|c|c|c|}
        \hline
        \textbf{$\epsilon$} & $10^{-2}$ & $10^{-4}$ & $10^{-6}$  \\
        \hline
        \textbf{$\sigma_c$} & 0.62667334 & 0.62666332 & 0.62661101  \\
        \hline
    \end{tabular}
    \caption{Estimation of the continuous transition point $\sigma_c$ in Fig.~\ref{fig:XY_Gaussian_ground_state_m_vs_sigma} as a function of the tolerance $\epsilon$, see text.}
    \label{tab:1}
\end{table}
A similar procedure, applied to determine the transition point $h_0^c$ in Fig.~\ref{fig:XY_bimodal_ground_m_vs_h0}, yields results reported in Table~\ref{tab:2}.    
\begin{table}[h]
    \centering
    \renewcommand{\arraystretch}{1.8} 
    \begin{tabular}{|c|c|c|c|}
        \hline
        \textbf{$\epsilon$} & $10^{-2}$ & $10^{-4}$ & $10^{-6}$  \\
        \hline
        \textbf{$h_0^c$} & 0.44220502 &  0.44220500  & 0.44220100  \\
        \hline
    \end{tabular}
    \caption{Estimation of the first-order transition point $h_0^c$ in Fig.~\ref{fig:XY_bimodal_ground_m_vs_h0}  as a function of the tolerance $\epsilon$, see text.}
    \label{tab:2}
\end{table}

We now proceed to estimate the tricritical point 
$(h_0^{\mathrm{tri}}, T^{\mathrm{tri}})$ shown in Fig.~\ref{fig:n2-bimodal-phase-diag}. 
To this end, starting with small values of $h_0$, we identify the largest value of $h_0$ 
for which a continuous phase transition is still observed, along with its corresponding 
critical temperature $T_c$. This pair $(h_0^{\mathrm{tri}}, T^{\mathrm{tri}})$ marks the tricritical point. Setting a tolerance $\epsilon$ as done above, the obtained results are shown in Table~\ref{tab:integrability_breaking}.
\begin{table}[h]
    \centering
    \renewcommand{\arraystretch}{1.8} 
    \begin{tabular}{|c|c|c|c|}
        \hline
        \textbf{$\epsilon$} & $10^{-2}$ & $10^{-4}$ & $10^{-6}$  \\
        \hline
        \textbf{$h_0^{\mathrm{tri}}$} & 0.35124007  &  0.35123860 &  0.35123775  \\
        \hline
        \textbf{$T_0^{\mathrm{tri}}$} & 0.30060055  & 0.30060050 & 0.30060050\\
        \hline
    \end{tabular}
    \caption{Estimation of the tricritical point in Fig.~\ref{fig:n2-bimodal-phase-diag} as a function of the tolerance $\epsilon$, see text.}
    \label{tab:integrability_breaking}
\end{table}

\clearpage

\bibliographystyle{unsrturl}
\bibliography{bibliography}

\end{document}